\documentclass[twocolumn,aps,showpacs,pra]{revtex4-2}
\usepackage{amsmath}
\usepackage{amssymb}
\usepackage{esint}
\usepackage{graphicx}

\usepackage[normalem]{ulem}
\usepackage[colorlinks=true,allcolors=blue,bookmarks=false,pdfusetitle]{hyperref}
\usepackage{url}

\makeatletter

\usepackage{color}

\usepackage{breakurl}

\begin{document}

\title{Frequency beating and damping of breathing  oscillations of a harmonically trapped one-dimensional quasicondensate}

\author{F. A. Bayocboc, Jr.}
\author{K. V. Kheruntsyan}
\affiliation{School of Mathematics and Physics, The University of Queensland, Brisbane, Queensland 4072, Australia}

\date{\today}

\begin{abstract}
We study the breathing (monopole) oscillations and their damping in a harmonically trapped one-dimensional (1D) Bose gas in the quasicondensate regime using a finite-temperature classical field approach. By characterising the oscillations via the dynamics of the density profile's rms width over long time, we find that the rms width displays beating of two distinct frequencies. This means that 1D Bose gas oscillates not at a single breathing mode frequency, as found in previous studies, but as a superposition of two distinct breathing modes, one oscillating at frequency close to $\simeq\!\sqrt{3}\omega$ and the other at $\simeq\!2\omega$, where $\omega$ is the trap frequency. The breathing mode at $\sim\!\sqrt{3}\omega$ dominates the beating at lower temperatures, deep in the quasicondensate regime, and can be attributed to the oscillations of the bulk of the density distribution comprised of particles populating low-energy, highly-occupied states. The breathing mode at $\simeq\!2\omega$, on the other hand, dominates the beating at higher temperatures, close to the nearly ideal, degenerate Bose gas regime, and is attributed to the oscillations of the tails of the density distribution comprised of thermal particles in higher energy states. The two breathing modes have distinct damping rates, with the damping rate of the bulk component being approximately four times larger than that of the tails component.
\end{abstract}

\maketitle

\section{Introduction}
\label{sec:breathing_intro}	

The study of low-energy excitations and their damping is an indispensable tool for the understanding of collective many-body effects in ultracold quantum gases. In particular, the temperature dependence of the frequency of collective oscillations and their damping have been the subject of scrutiny both experimentally \cite{jin1996,jin1997,stamper-kurn1998,marago2001,chevy2002}  and theoretically \cite{fedichev1998,pitaevskii1997,pitaevskii1998,bijlsma1999,alKhawaja2000,guilleumas1999,jackson2001,jackson2002a,jackson2002b,jackson2002c,jackson2003,guilleumas2003} since the first experiments in dilute gas Bose-Einstein condensates \cite{anderson1995,bradley1995,davis1995b}. Depending on the temperature of the gas, the damping of collective oscillations in harmonically trapped 3D systems has been explained either via collisional relaxation \cite{jackson2002a,griffin_nikuni_zaremba_2009}, where the two parts of the Bose gas (condensate and thermal components) exchange energy and particles, or via mean-field effects that can lead to Landau or Beliaev mechanisms of damping \cite{pitaevskii1997,fedichev1998,guilleumas1999,jackson2003,beliaev1958}. The lifetime of collective oscillations in such systems has been predicted and measured to be typically on the order of tens of milliseconds.

In contrast to 3D systems, collective oscillations in one-dimensional (1D) Bose gases damp out on a significantly longer time scales. For example, the lifetime of breathing mode oscillations observed in Ref. \cite{fang2014} in a weakly interacting 1D quasicondensate was on the order of seconds; in the related collisional dynamics of a quantum Newton's cradle in the opposite, strongly interacting regime, the thermalisation time constant was estimated to be even longer (longer than $\sim$70 seconds) \cite{kinoshita2006}. The slow relaxation rates in the 1D Bose gas are related to the fact that the underlying theoretical model---the Lieb-Liniger model \cite{lieb1963a,lieb1963b}---is integrable in the uniform limit, which puts additional constraints on the pathways to equilibration compared to those present in generic (non-integrable) quantum systems. More specifically, the integrable uniform 1D Bose gas is expected to relax to a generalised Gibbs ensemble rather than to the canonical thermal state \cite{Rigol2008,Rigol2009,polkovnikov2011,Kollar2011,Caux2012,gring2012,langen2015b}. In inhomogeneous 1D Bose gases, such as the harmonically trapped 1D quasicondensate studied here, the integrability breaks down and provides a mechanism for relaxation to a thermal ensemble \cite{mazets2011}. Nevertheless, for sufficiently weak confinement, the system can be regarded as nearly-integrable and hence is expected to undergo a crossover from transient relaxation to the generalised Gibbs state to a slow decay to the final thermal ensemble \cite{gring2012}. The overall 1D damping rate is expected to be small enough to be neglected in experiments. However, in current experiments the observed relaxation rates in quasi-1D systems are often affected by transverse excitations \cite{mazets2008,mazets2010,tan2010,GHD_extension} due to the 3D nature of realistic trapping potentials. Such transverse excitations speed up thermalisation, thus hampering the characterisation of pure 1D damping. Because of this, pure 1D damping rates have not been scrutinised experimentally yet, particularly in the weakly interacting, degenerate regime of the 1D Bose gas, whereas theoretically the question of 1D thermalisation has started to attract attention only relatively recently \cite{Bland_2018,bastianello2020,Thomas_2021}. 

In this paper, we study damping rates of a finite-temperature 1D Bose gas in the weakly interacting quasicondensate regime, following an excitation of breathing mode oscillations in a harmonic trap. The specific scenario that we consider is a sudden trap quench from the initial trap frequency $\omega_0$ to a slightly smaller frequency $\omega$, which invokes breathing oscillations; we simulate these oscillations and their relaxation dynamics using a classical field ($c$-field) approach. In doing so, we also revisit and scrutinise the question of the frequency of breathing oscillations, which has been addressed previously both experimentally \cite{moritz2003,haller2009,fang2014} and theoretically \cite{Sinatra_Lobo_Castin_2001,menotti2002,pedri2003,Schmitz2013,Haque2013,Reentrant2015,Olshani_monopole_2015,Hu2015,DeRosi2015,DeRosi2016,bouchoule2016}.

According to the most recent experimental study by Fang \emph{et al.} \cite{fang2014}, the frequency of such oscillations in the root-mean-square (rms) width of the real-space density profile undergoes a smooth transition from $\omega_B\simeq \sqrt{3}\omega$ deep in the quasicondensate regime to $\omega_B \simeq 2\omega$ in the nearly ideal Bose gas regime as the temperature of the gas is increased. In contrast to this, our numerical experiment reveals the presence of both oscillation frequencies in a broad range of temperatures within the quasicondensate regime. We refer to these frequencies as $\omega_{B1}$ and $\omega_{B2}$ and attribute the breathing modes at $\omega_{B1}\simeq \sqrt{3}\omega$ and $\omega_{B2}\simeq 2\omega$, respectively, to the oscillations of the bulk and the tail components of the density profile. The observation of two simultaneous breathing modes is made possible by extending our dynamical simulations to significantly longer durations than currently possible experimentally, which reveals an oscillatory pattern (in the rms width) characteristic of beating of two frequencies. Such beating in breathing oscillations, resulting in `collapses' and `revivals' of the rms width of the density profile, is similar to the one observed recently in a partially condensed 3D Bose-Einstein condensate \cite{straatsma2016}.

Finally, we extract the damping rates of these two distinct breathing modes from the $c$-field simulations and find that the damping rate $\Gamma_1$ associated with the breathing mode  $\omega_{B1}$ is on the order of $\Gamma_1 \simeq 0.04 \omega$ (where we use $1/\omega$ as the timescale), whereas the damping rate $\Gamma_2$ associated with $\omega_{B2}$ is approximately four times smaller. 

\section{$C$-field method for simulating breathing oscillations}
\label{Method}

The breathing mode oscillations of a 1D Bose gas in the quasicondensate regime are simulated using the $c$-field (or classical field) approach as in Refs.~\cite{bouchoule2016,Thomas_2021,bayocboc2021dynamics}. In this approach \cite{castin2000,blakie2008}, the initial thermal equilibrium state of the system is prepared by evolving the simple
growth stochastic projected Gross-Pitaevskii equation
(SPGPE) for the complex $c$-field $\Psi_{\mathbf{C}}(x,t)$,
\begin{align}
	\label{eq:breathing_SPGPE}
	\mathrm{d}&\Psi_{\mathbf{C}}(x,t)=\mathcal{P}^{(\mathbf{C})}\!\left\{ - \frac{i}{\hbar}\mathcal{L}_{0}^{(\mathbf{C})}\Psi_{\mathbf{C}}(x,t)\,\mathrm{d}t \right. \nonumber\\
	&\left.+ \frac{\Gamma}{k_BT}(\mu-\mathcal{L}_{0}^{(\mathbf{C})})\Psi_{\mathbf{C}}(x,t)\, \mathrm{d}t + dW_{\Gamma}(x,t)\!\right\},
\end{align}
with $x$ and $t$ being the position and time, respectively. Here, the projection operator $\mathcal{P}^{(\mathbf{C})}\{\cdot\}$ sets up the high-energy cutoff $E_{\mathrm{cut}}$
between the classical $c$-field region, comprised of degenerate, highly occupied low-energy modes, and the thermal region, comprised of  sparsely occupied high-energy modes. Furthermore, $\Gamma$ is the growth rate, $T$ is the temperature of the effective reservoir (served by the thermal region) to which the system is coupled, and $\mu$ is the chemical potential of the reservoir that controls the number of particles in the \textit{c}-field region. In addition, $\mathcal{L}_{0}^{(\mathbf{C})}$ is the Gross-Pitaevskii operator defined by
\begin{equation}
	\mathcal{L}_{0}^{(\mathbf{C})} = -\frac{\hbar^{2}}{2m}\frac{\partial^{2}}{\partial x^{2}} + V(x,t) + g|\Psi_{\mathbf{C}}(x,t)|^{2}, 
\end{equation}
where $V(x,t)$ is the external trapping potential, which we assume is harmonic, $V(x,t)=\frac{1}{2}m\omega(t)^2x^2$, with frequency $\omega(t)$, and $g$ is the strength of repulsive ($g>0$) 1D interatomic contact interaction potential. The last term, $dW_{\Gamma}(x,t)$, in Eq.~\eqref{eq:breathing_SPGPE} is a complex-valued stochastic white noise satisfying the following nonzero correlation:
\begin{equation}
	\langle dW_{\Gamma}^{*}(x,t)dW_{\Gamma}(x',t) \rangle = 2\Gamma\delta(x-x')dt.
\end{equation}

Evolving the above SPGPE from an arbitrary initial state and for sufficiently long time (such that the memory of the initial state is lost) samples thermal equilibrium configurations of the system from the grand-canonical ensemble. Averages over a large number of stochastic realisations of the $c$-field $\Psi_{\mathbf{C}}(x,t)$ and its complex conjugate $\Psi_{\mathbf{C}}^{*}(x,t)$ are then used to construct thermal equilibrium values of physical observables that can be expressed in terms of expectation values of standard bosonic quantum field operators $\hat{\Psi}(x,t)$ and $\hat{\Psi}^{\dagger}(x,t)$, except that their quantum commutating nature is ignored. As an example, the particle number density $\rho(x,t) = \langle \hat{\Psi}^{\dagger}(x,t) \hat{\Psi}(x,t) \rangle$ in the $c$-field approach is calculated as the stochastic average $\rho(x,t)  = \langle \Psi_{\mathbf{C}}^{*}(x,t) \Psi_{\mathbf{C}}(x,t) \rangle$ (where the brackets $\langle{...}\rangle$ now refer to stochastic averaging over a large number of stochastic trajectories), whereas the momentum distribution $n(k,t)$, where $k$ is in wave-number units, is calculated as $n_j(k,t)\!=\!\iint dx \,dx'e^{ik(x-x')} \langle \Psi_{\mathbf{C}}^{*}(x,t) \Psi_{\mathbf{C}}(x',t)  \rangle$.

The thermal equilibrium configurations (stochastic realisations) of the $c$-field $\Psi_{\mathbf{C}}(x,t)$, prepared in this way via the SPGPE, form the initial ($t=0$) thermal equilibrium state of the system. The subsequent dynamics of the system, following a certain excitation protocol, can then be modelled by evolving the $c$-field realisations in real time according to the mean-field projected Gross-Pitaevskii equation \cite{blakie2008},
\begin{equation}
	i\hbar\frac{\partial}{\partial t}\Psi_{\mathbf{C}}(x,t) =\mathcal{P}^{(\mathbf{C})}\!\left\{  \mathcal{L}_{0}^{(\mathbf{C})}  \Psi_{\mathbf{C}}(x,t) . \right\},
	\label{GPE}
\end{equation}

The dynamical protocol that we use here to invoke the breathing mode oscillations is a sudden quench (at time $t=0$) of the harmonic  trap frequency from $\omega_0$ to a new value $\omega$, i.e.,
\begin{equation}
V(x,t)=\begin{cases}
\frac{1}{2}m\omega_{0}^{2}x^{2}, & \mathrm{{for}}\,\,t\leq0,\\
\frac{1}{2}m\omega^{2}x^{2}, & \mathrm{{for}}\,\,t>0.
\label{sudden}
\end{cases}
\end{equation}

The strength of such a quench can be characterised by
\begin{equation}
	\epsilon = \left( \frac{\omega_{0}}{\omega} \right)^{2} - 1,
\end{equation}
which can be either negative or positive depending on the ratio $\omega_0/\omega$ being smaller or larger than one. The numerical value of $\epsilon$ determines the amplitude of breathing mode oscillations; for a small-amplitude quench, with $|\epsilon|\ll 1$, the amplitude of oscillations is linear in $\epsilon$, according to the scaling solutions to hydrodynamic equations of Ref.~\cite{bouchoule2016} (see footnote [33] therein).

\begin{figure*}[tbp]
		\includegraphics[width=17cm]{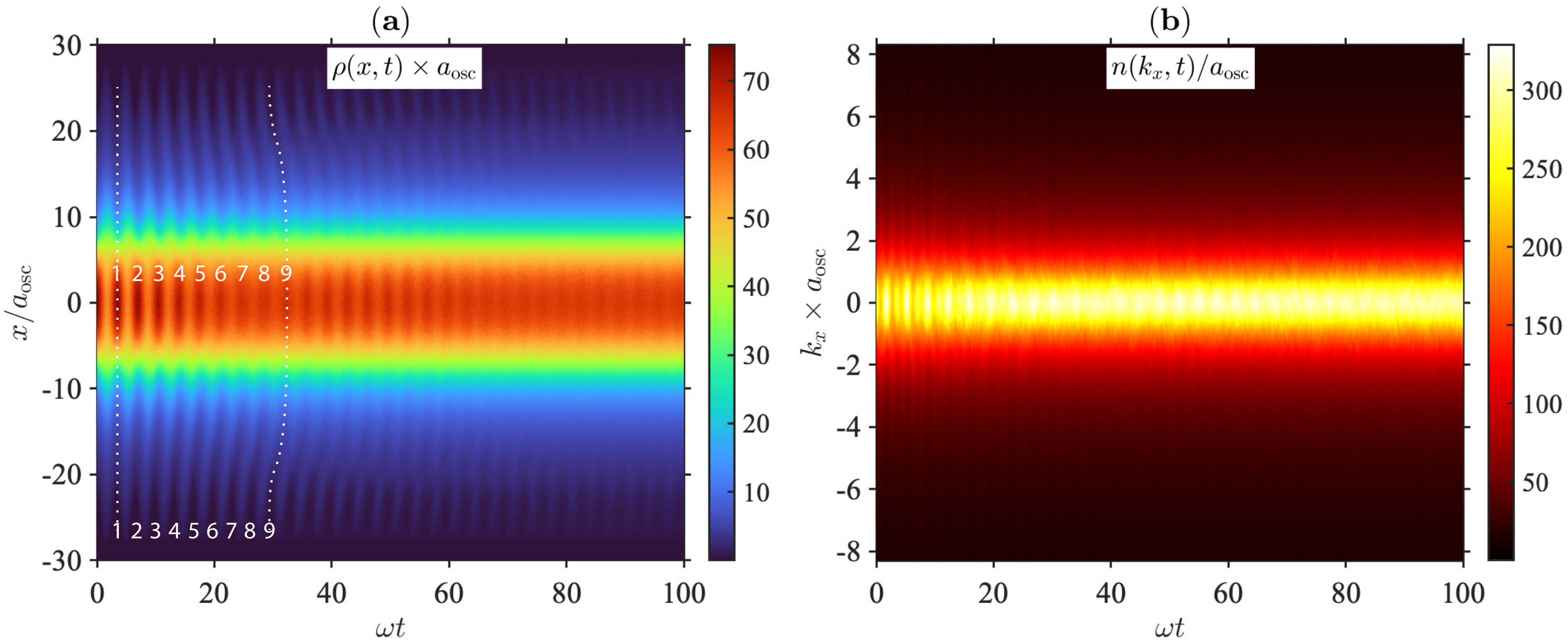}
		\caption{Typical evolution of the real-space density profile $\rho(x,t)$ and the momentum distribution $n(k_{x},t)$ of a quasicondensate after a quench of the longitudinal trapping frequency $\omega_{0} \to \omega$, with a quench strength $\epsilon \simeq 0.563$ ($\omega/\omega_0=0.8$). The initial thermal equilibrium state of the system is characterised by $\gamma_{0}^{3/2}\mathcal{T} = 0.39$ and $N=1109$. The dimensionless position ($x/a_{\mathrm{osc}}$) and momentum ($ka_{\mathrm{osc}}$) are introduced with respect to the initial harmonic oscillator length $a_{\mathrm{osc}}=\sqrt{\hbar/m\omega_0}$ serving as the lengthscale, whereas the time is normalised to $1/\omega$. In terms of absolute values, the relevant physical parameters here were chosen as follows: $T=48$ nK, $\omega_0/2\pi=10$ Hz, and $g\simeq1.4 \times 10^{-38}$ J$\cdot$m, which itself can be obtained from $g\simeq 2\hbar\omega_{\perp}a$ \cite{olshanii1998}, with the 3D scattering length of $a=5.3$ nm (assuming a gas of $^{87}$Rb atoms) and the frequency of transverse harmonic confinement of $\omega_{\perp}/2\pi=2$ kHz (with $g\simeq2\hbar\omega_{\perp}a$ away from confinement induced resonances \cite{olshanii1998}). With these parameter values, one obtains $\gamma_0=8.3 \times 10^{-3}$, $\mathcal{T}=517$, and the resulting quoted values of $\gamma_{0}^{3/2}\mathcal{T} = 0.39$ and $N=1109$. Here and hereafter, all observables are evaluated as averages over $2000$ stochastic realisations. The dotted vertical lines in (a) are a guide to an eye showing that the oscillation frequency in the density profile varies depending on the position $|x|$ from the trap centre (see text).}
		\label{fig:posdens_momdens_sample}
\end{figure*}

Breathing oscillations of a 1D quasicondensate in this particular scenario have been studied previously experimentally and theoretically in Refs.~\cite{fang2014,bouchoule2016}. The focus of those works was the understanding of the phenomenon of frequency doubling of the oscillations in momentum space. In the present work, we instead concentrate on analysing the frequencies and damping rates of breathing oscillations seen in Refs.~\cite{fang2014,bouchoule2016}, in analogy with a recent work on thermalisation of a 1D quasicondensate in a quantum Newton's cradle setup \cite{Thomas_2021}.

The simulations of the SPGPE and projected GPE performed in this work were carried out using the XMDS software package \cite{XMDS} in the computational basis of Hermite-Gauss polynomials. Unlike the plane-wave basis, which necessitates additional optimisation of the computational grid to control the effects of the high-energy cutoff \cite{Bradley_2005}, the Hermite-Gauss basis is the natural basis for harmonically trapped systems and represents the most computationally efficient basis \cite{blakie2008,Blakie_2008_Numerical_Method,rooney2014}. In this basis, the projection operator $\mathcal{P}^{(\mathbf{C})}\{\cdot\}$ that provides the high-energy cutoff, with $E_{\mathrm{cut}}=\hbar\omega_0(n_{\max}+1/2)$, is implemented naturally through the maximum number of basis states $n_{\max}$ used in the numerical simulations.

Before discussing the results of our simulations, we refer the reader to Appendices \ref{appendix:Regimes}, \ref{c-field-regime} and \ref{trapped}, where we recall the classification of different regimes of a weakly interacting uniform 1D Bose gas, the regimes of applicability of the $c$-field approach, and their extensions to a harmonically trapped 1D Bose gas. The different regimes here are identified via two dimensional parameters: $\gamma_0=mg/\hbar^2\rho_0$, characterising the interaction strength for a uniform system at density $\rho_0$, and $\mathcal{T} \!=\! 2\hbar^{2}k_{\mathrm{B}}T/mg^{2}$, characterising the temperature. Furthermore, the $c$-field approach can, in fact, be parametrised in terms of a single dimensionless parameter, $\gamma_{0}^{3/2}\mathcal{T}$ \cite{castin2000,bouchoule2012,bouchoule2016,Thomas_2021}. For a harmonically trapped (inhomogeneous) system, the roles of $\rho_0$ and $\gamma_0$ are taken by their respective values in the trap centre, with $\gamma_{0}^{3/2}\mathcal{T}$ still being a relevant single dimensionless parameter that combines interaction strength and tempeprature. However, to completely characterise such a system, one needs to specify an additional parameter, such as the total number of atoms $N$, which itself is governed (for fixed values of $\gamma_0$ and $\mathcal{T}$)  by the underlying trap frequency $\omega_0$.

In Fig.~\ref{fig:posdens_momdens_sample}, we show typical evolution of the density profile $\rho(x,t)$ and the respective momentum distribution $n(k_{x},t)$, after a sudden quench of the trap frequency as described above. In this example, the initial state is characterised by $\gamma_{0}^{3/2}\mathcal{T} = 0.39$ and a total number of atoms $N=1109$, whereas the quench strength is $\epsilon = 0.563$ ($\omega/\omega_0=0.8)$. We see here that both $\rho(x,t)$ and $n(k_{x},t)$ display breathing oscillations after the quench, with $n(k_{x},t)$ showing additional peaks occurring at twice the frequency of oscillations of $\rho(x,t)$. This phenomenon is known as frequency doubling \cite{fang2014} and can be interpreted, via a classical hydrodynamic approach \cite{bouchoule2016} as a result of a self-reflection mechanism at the inner turning point due to the mean-field interaction energy barrier. Similarly to the results of Ref.~\cite{bouchoule2016}, we observe the frequency doubling in this example because the system under these parameters is the regime where the contribution of the hydrodynamic velocity field dominates the contribution of the thermal velocities (which show no frequency doubling). As the system evolves in time the oscillations in both the density and momentum distributions can be seen to damp out, with the damping somewhat more apparent in the momentum distribution.

 The dotted vertical lines in Fig.~\ref{fig:posdens_momdens_sample}\,(a) are shown to guide the eye in the observation that the oscillation frequency varies depending on the position $|x|$; the frequency of local density oscillations near the trap centre is seen to be lower than the local frequency in the tails of the density distribution. Indeed, while the first density maximum in the trap centre and the respective minimum in the tails are aligned to a straight vertical line, the ninth maximum in the trap centre is ahead of the respective ninth minimum in the tails, implying that the central part of the cloud oscillates at a lower frequency.

\section{Breathing dynamics}

To further characterise the dynamics and damping of the breathing oscillations, we calculate the rms width of the density profile, given by
\begin{equation} \label{eq:RMS_width}
	\Delta x_{\mathrm{RMS}}(t) \!=\! \left[ \frac{1}{N}\! \int \!\mathrm{d}x \rho(x,t)x^{2} \!-\! \Bigl( \frac{1}{N}\! \int \!\mathrm{d}x \rho(x,t)x \Bigr)^{2} \right]^{1/2},
\end{equation}
where $N = \int\mathrm{d}x\rho(x)$ is the total number of particles.

\begin{figure}[tbp]
	\begin{center}
		\includegraphics[width=0.45\textwidth]{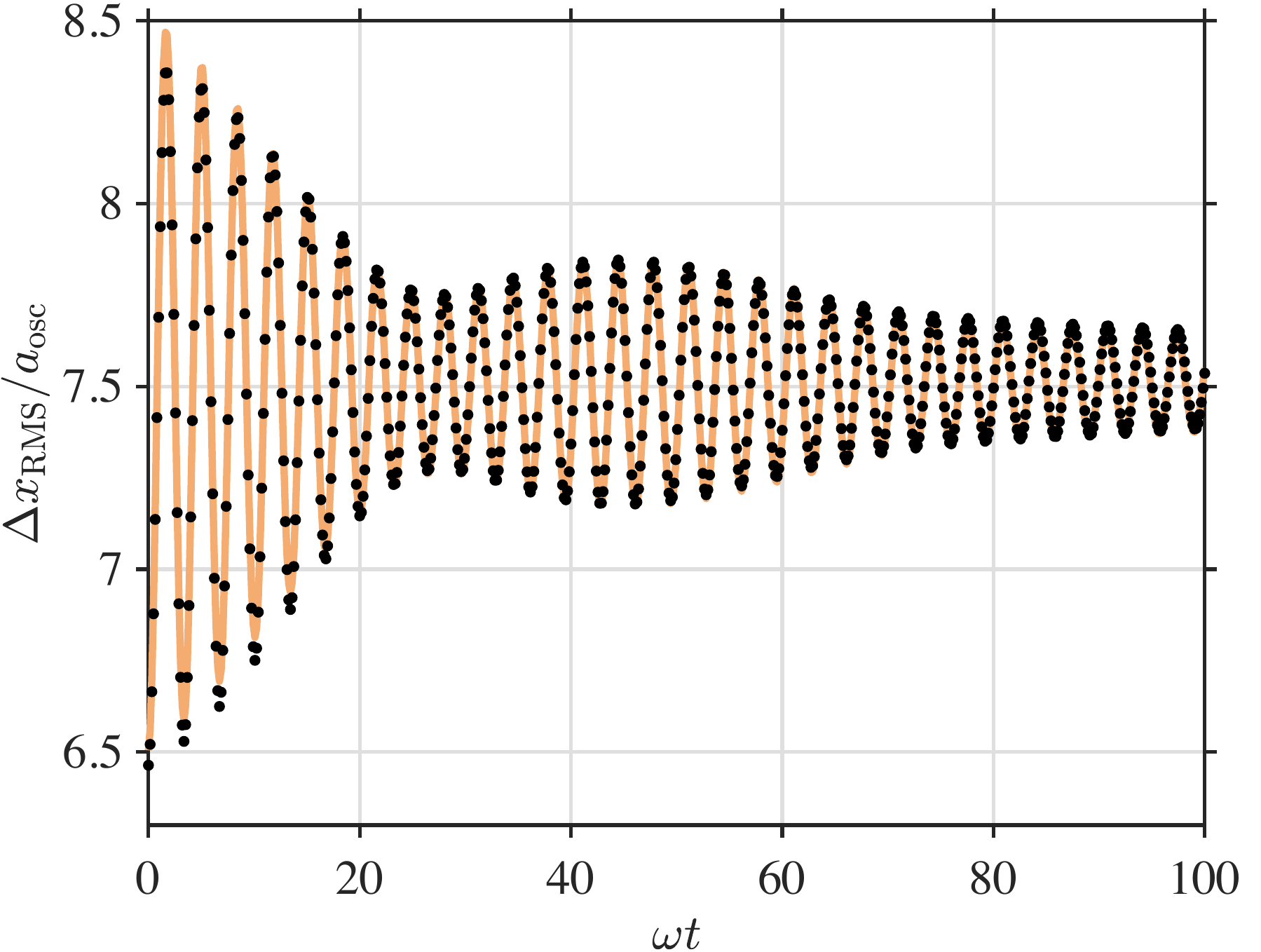}
		\caption{Root-mean-square width $\Delta x_{\mathrm{RMS}}(t)$ of the density profile $\rho(x,t)$ shown in Fig.~\ref{fig:posdens_momdens_sample}\,(a), normalised to harmonic oscillator length $a_{\mathrm{osc}}=\sqrt{\hbar/m\omega_0}$. The black dots are data points from $c$-field simulations, whereas the full (orange) line is a fit using Eq.~\eqref{eq:RMS_fit}.}
		\label{fig:RMS_sample}
	\end{center}
\end{figure}

In Fig.~\ref{fig:RMS_sample}, we show the calculated $\Delta x_{\mathrm{RMS}}(t)$ as a function of time, for the density profile $\rho(x,t)$ of Fig.~\ref{fig:posdens_momdens_sample}\,(a). A distinct feature of the rms width is the presence of beating of two oscillation frequencies, which was already apparent in Fig.~\ref{fig:posdens_momdens_sample}\,(a). This beating suggests that the quasicondensate, after the quench of the trapping potential, oscillates not at a single breathing mode frequency, but as a superposition of two dominant frequencies.

To extract the beating frequencies from the oscillations of the rms width, we fit a sum of two cosine functions, each with its own  damping term,
\begin{align} \label{eq:RMS_fit}
	\Delta x_{\mathrm{RMS}}(t) &= A_{1}\cos(\omega_{\mathrm{B}1}t + \phi_{1})e^{-\Gamma_{1}t} \nonumber \\
	&+ A_{2}\cos(\omega_{\mathrm{B}2}t + \phi_{2})e^{-\Gamma_{2}t} + C.
\end{align}
Here, $\omega_{Bi}$ ($i=1,2$) are the two breathing mode frequencies, $A_i$, $\Gamma_{i}$ and $\phi_{i}$ are the respective amplitudes, damping rates, and the phases 
of each breathing mode, and the last term $C$ serves as a constant background. As can be seen from Fig.~\ref{fig:RMS_sample}, Eq.~\eqref{eq:RMS_fit} fits very well to the rms width $\Delta x_{\mathrm{RMS}}(t)$ calculated from the \textit{c}-field simulations, confirming that the nontrivial oscillatory dynamics of the rms width $\Delta x_{\mathrm{RMS}}(t)$ is indeed a result of beating of two components of the 1D Bose gas, breathing at two distinct frequencies $\omega_{B1}$ and $\omega_{B2}$.

Similar beating in the rms width has been observed in 3D systems \cite{straatsma2016}, where the effect was referred to as `collapses' and `revivals' of the rms width  due to in-phase and out-of-phase oscillations of the condensed and noncondensed fractions of the gas.

We further note that the same fitting formula, Eq.~\eqref{eq:RMS_fit}, works very well when applied to the rms width of the density distribution evaluated for single stochastic realisations of the SPGPE and projected GPE, albeit with slightly different (fluctuating) values of the oscillation amplitudes $A_{1,2}$ and the background constant $C$. This implies that the decay of oscillations in time is indeed consistent with exponential damping, rather than is a seeming decay due to blurring or averaging over many stochastic realisations with varying relative phase offsets.

\subsection{Beating frequencies}
\label{beating_frequencies}

To understand the emergence of two distinct frequencies in the breathing oscillations of a 1D quasicondensate, we recall the results of Ref.~\cite{bouchoule2016}, in which the breathing dynamics were studied within the classical hydrodynamic approach. The relevant regimes considered in Ref.~\cite{bouchoule2016} were: a weakly interacting 1D Bose gas deep in the quasicondensate regime, $\gamma_0^{-1}\ll \mathcal{T}\ll \gamma_0^{-3/2}$ (or $g\rho_0\ll k_{\mathrm{B}}T\ll \sqrt{\gamma_0}\hbar^2\rho_0^2/2m$), characterised by suppressed density fluctuations and a fluctuating phase; and (b) 
nearly ideal but highly degenerate 1D Bose gas, $\gamma_0^{-3/2} \ll \mathcal{T}\ll \gamma_0^{-2}$ (or $\sqrt{\gamma_0}\hbar^2\rho_0^2/2m \ll k_{\mathrm{B}}T \ll \hbar^2\rho_0^2/2m$), in which both the density and phase fluctuate (see also Appendices \ref{appendix:Regimes}, \ref{c-field-regime} and \ref{trapped}).

According to Refs.~\cite{bouchoule2016}, the frequencies of breathing mode oscillations in the quasicondensate and nearly ideal degenerate Bose gas regimes, found from hydrodynamic scaling solutions, are given by $\omega_{B}\!=\!\sqrt{3}\omega$ and $\omega_{B}\!=\!2\omega$, respectively. We further note here, that the breathing frequency $\omega_{B}\!=\!2\omega$ extends into the nearly ideal but nondegenerate (classical) gas regime \cite{DeRosi2015,DeRosi2016,bouchoule2016}, corresponding to $\mathcal{T}\gg \gamma_0^{-2}$, however, this regime is beyond the applicability of the $c$-field approach (see Appendices \ref{c-field-regime} and \ref{trapped}) that we use in this work.

Other theoretical and experimental studies of harmonically trapped 1D Bose gas \cite{menotti2002,moritz2003,haller2009,Schmitz2013,Haque2013,fang2014,Reentrant2015,Olshani_monopole_2015,Hu2015,DeRosi2015,DeRosi2016}, have predicted and observed breathing mode oscillation frequencies close to $2\omega$ and $\sqrt{3}\omega$. Furthermore, for a zero temperature gas, the breathing mode dynamics was predicted \cite{Reentrant2015,Hu2015} to display the so-called reentrant behaviour, wherein the frequency of oscillations was shown to undergo a smooth crossover from the ideal Bose gas value of $2\omega$ down to $\sqrt{3}\omega$ in the weakly interacting gas, and then back to $2\omega$ in the strongly interacting regime.

\begin{figure}[tbp]
	\begin{center}
		\includegraphics[width=0.45\textwidth]{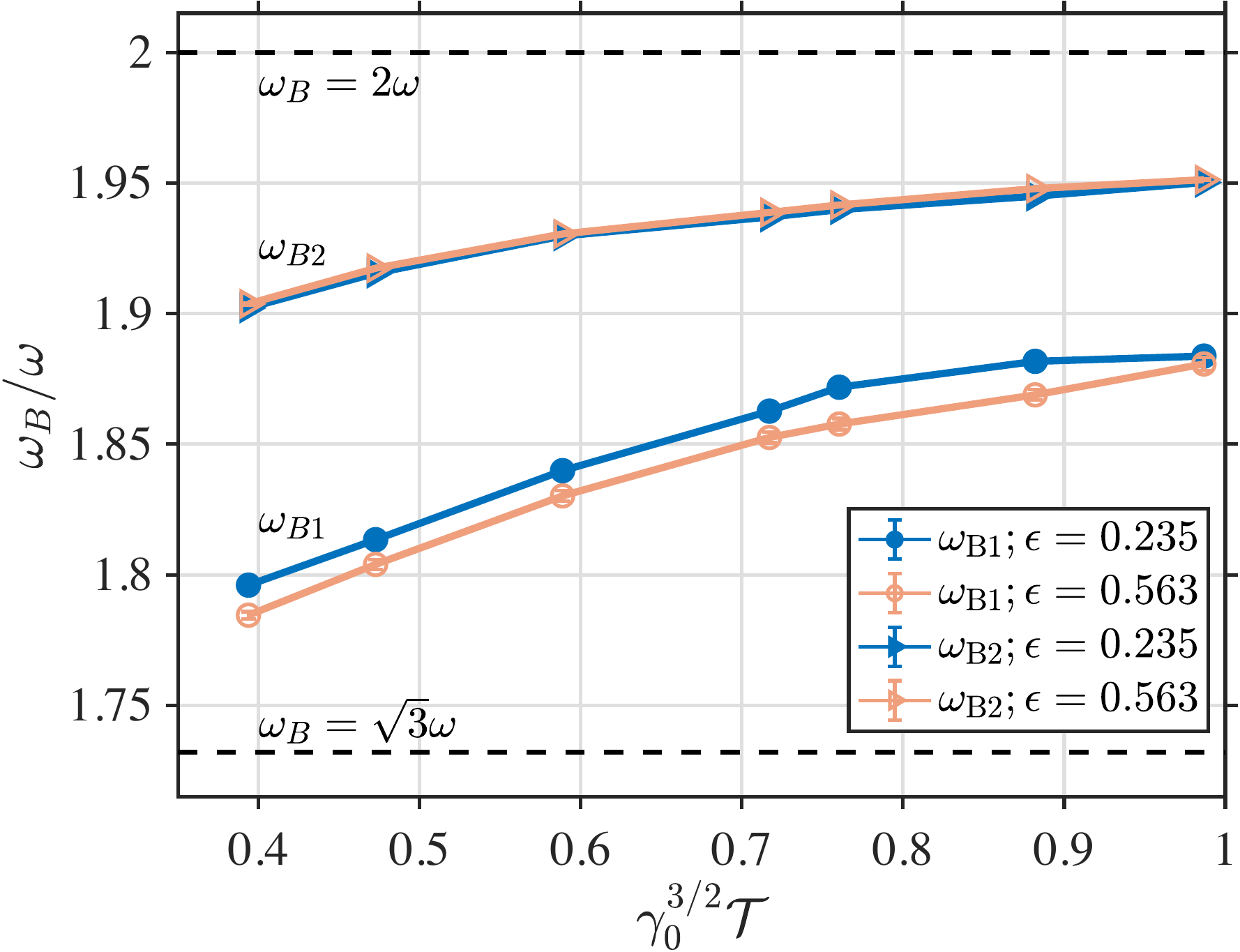}
		\caption{Breathing mode frequencies $\omega_{\mathrm{B}1}$ and $\omega_{\mathrm{B}2}$ as a function of  $\gamma_{0}^{3/2}\mathcal{T}$ extracted from $c$-field simulations by using the fitting equation Eq.~\eqref{eq:RMS_fit}. The error bars account for fitting error only and indicate a 95\% confidence interval. Two different sets of data points correspond to two values of the trap quench strength $\epsilon$: $\epsilon \simeq 0.235$ ($\omega/\omega_0=0.9$) and $\epsilon \simeq 0.563$ ($\omega/\omega_0=0.8$). The first data point here, which is for the smallest value of $\gamma_{0}^{3/2}\mathcal{T}=0.39$ and $N=1109$, is the same as in Fig.~\ref{fig:posdens_momdens_sample}. The subsequent, increasingly larger values of $\gamma_{0}^{3/2}\mathcal{T}$ were achieved by keeping the temperature $T$ and the initial trap frequency $\omega_0$ the same and scanning the chemical potential of the system, which then determines the total number of atoms $N$ (as well as the peak density $\rho_0$ and $\gamma_0$). The respective values of $N$ as a function of $\gamma_{0}^{3/2}\mathcal{T}$ for all data points are shown in Fig.~\ref{fig:condensate_fraction} below.
		}
		\label{fig:fit_omegaB}
	\end{center}
\end{figure}

However, as we have seen from our $c$-field simulations of the previous section, the breathing oscillations in a finite temperature quasicondensate display a beating of two distinct frequencies. This suggests that in a weakly interacting 1D Bose gas, the bulk of the quasicondensate density near the trap centre, where the interactions are more important, oscillates at the frequency close to $\sqrt{3}\omega$, whereas the tails of the density distribution, behaving more like an ideal (noninteracting) Bose gas, oscillate at the frequency closer to $2\omega$. To confirm this hypothesis, we now simulate the dynamics of breathing oscillations for different values of the dimensionless parameter $\gamma_{0}^{3/2}\mathcal{T}$, varying it in the range $0.39\!\leq \! \gamma_{0}^{3/2}\mathcal{T}\!\leq\!1$ (with $\mathcal{T}= 517$, $\gamma_0\!=\! 8.3\times 10^{-3}$, and $N=1109$ at the lower bound, and $\mathcal{T}= 517$, $\gamma_0\!=\! 1.5\times 10^{-2}$, and $N=609$ at the upper bound). This scans the conditions of our system in the trap centre from the thermal quasicondensate regime towards the crossover boundary with the nearly ideal degenerate Bose gas. Upon doing so, we extract the breathing mode frequencies $\omega_{Bi}$  ($i\!=\!1,2$) and the respective damping rates $\Gamma_i$ by fitting the rms width of the density distribution to Eq.~\eqref{eq:RMS_fit} for each value of $\gamma_{0}^{3/2}\mathcal{T}$.

In Fig.~\ref{fig:fit_omegaB}, we show the extracted frequencies as a function of $\gamma_{0}^{3/2}\mathcal{T}$, for two different values of the quench strength $\epsilon$. As we see, in both cases, and for the smallest values of $\gamma_{0}^{3/2}\mathcal{T}$, the extracted breathing mode frequency $\omega_{B1}$ is indeed close to the value of $\omega_{B1}\!= \!\sqrt{3}\omega$, whereas $\omega_{B2}$ is closer to $\omega_{B2}\!= \!2\omega$. As the dimensionless parameter $\gamma_{0}^{3/2}\mathcal{T}$ is increased towards the degenerate ideal Bose gas regime, $\gamma_{0}^{3/2}\mathcal{T}\!\simeq \!1$, the frequencies of both components increase too, with $\omega_{B1}$ deviating further away from the value of $\sqrt{3}\omega$ and both $\omega_{B1}$ and $\omega_{B2}$ tending towards $2\omega$.

\begin{figure}[tbp]
	\begin{center}
		\includegraphics[width=0.45\textwidth]{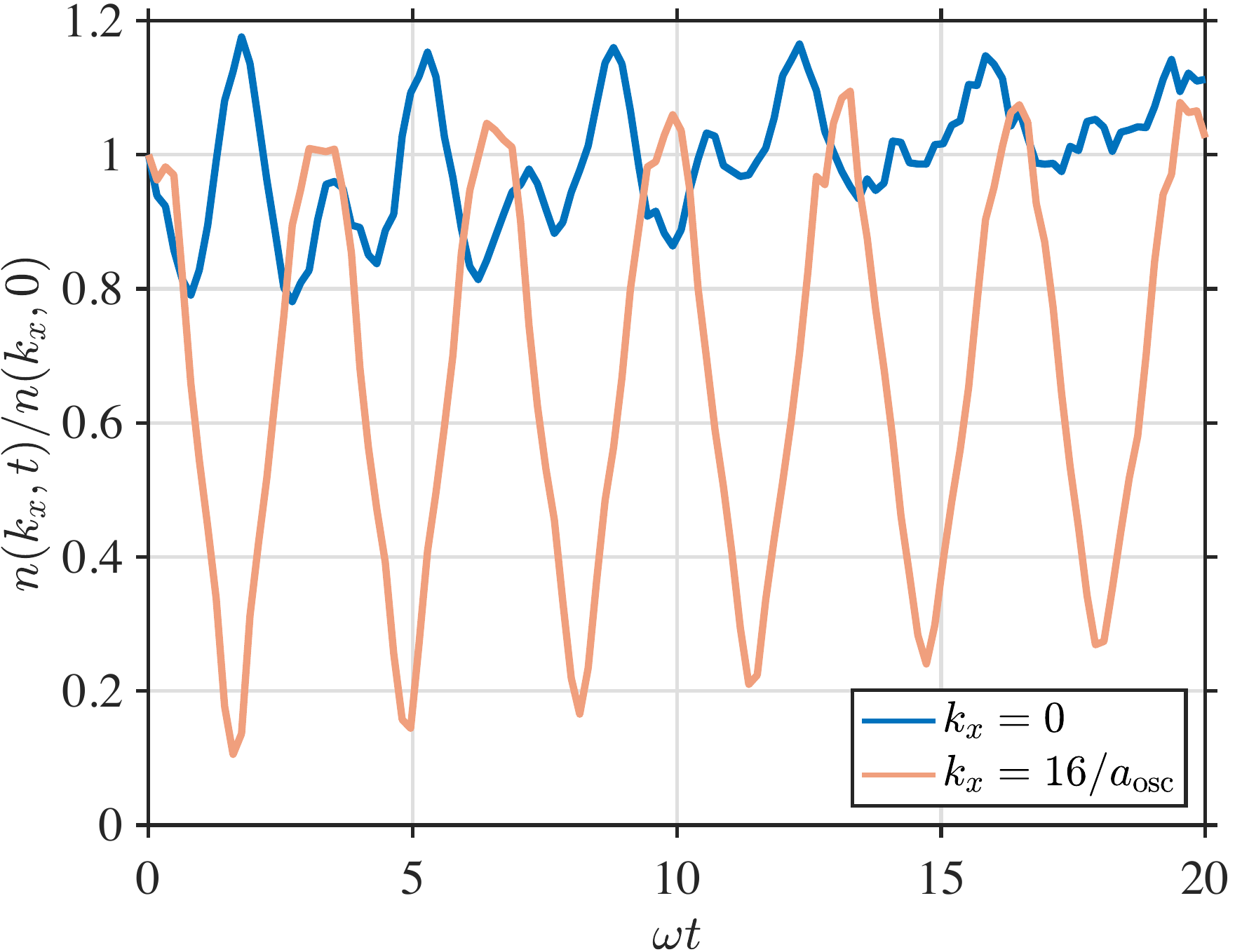}
		\caption{Evolution of the momentum distribution $n(k_{x},t)$ at $k_{x} = 0$ and at $k_{x} = 16/a_{\mathrm{osc}}$, for $\gamma_{0}^{3/2}\mathcal{T} = 0.39$, $N=1109$, and $\epsilon = 0.563$ ($\omega/\omega_0=0.8$). The momentum distribution at $k_{x} = 0$ displays frequency doubling, with $\omega^{(k)}\!=\!2\omega_{B1}$, which is characteristic of a system deep in the quasicondensate regime. On the other hand, the momentum distribution at $k_{x} = 16/a_{\mathrm{osc}}$ does not display frequency doubling and oscillates at $\omega^{(k)}\!= \!\omega_{B2}$, which is characteristic of an ideal Bose gas regime.
		}
		\label{fig:breathing_bulk_vs_tail}
	\end{center}
\end{figure}

One can identify the frequencies $\omega_{B1}$ and $\omega_{B2}$, respectively, with the breathing mode oscillations of the bulk of the quasicondensate near the trap centre (dominated by higher occupancy, low-energy states) and the tails of
the density profile (dominated by lower occupancy, high-energy
states). Indeed, particles near the trap centre have the local value of $\gamma_{x}^{3/2}\mathcal{T}\!< \!1$ (with $\gamma_{x}\! =\! mg/\hbar^{2}\rho(x)$) and hence are deeper in the quasicondensate regime, whereas particles in the tails of the density distribution have the local value of $\gamma_{x}^{3/2}\mathcal{T}\!> \!1$ and can be approximated as an ideal degenerate Bose gas.

This conclusion can be further verified if we inspect the local dynamics of the momentum distribution of the gas, $n(k_x,t)$. 
In Fig.~\ref{fig:breathing_bulk_vs_tail}, we can see the difference in the behaviour of the momentum distribution at $k_{x} = 0$ and at $k_{x} \!= \!16/a_{\mathrm{osc}}$. While the oscillations at $k_{x} \!=\! 0$ display frequency doubling ($\omega^{(k)}\!=\!2\omega_{B1}$) \cite{fang2014,bouchoule2016}, which is a property of a 1D Bose gas deep in the quasicondensate regime, the momentum distribution at $k_{x} \!=\! 16/a_{\mathrm{osc}}$ does not display frequency doubling and oscillates at frequency $\omega^{(k)}$ close to $\omega_{B2}$. The oscillation frequencies extracted from the two curves in Fig.~\ref{fig:breathing_bulk_vs_tail} yield $\omega_{B1}\! \approx \!1.79\omega$ for $n(k_{x}\!=\!0,t)$ and $\omega_{B2} \!\approx \!1.92\omega$ for $n(k_{x} \!=\! 16/a_{\mathrm{osc}},t)$. These frequencies are close to the breathing mode frequencies in Fig.~\ref{fig:fit_omegaB}, extracted from the rms width of the respective density distribution. Thus, from this point onwards, we will refer to the component with breathing frequency $\omega_{B1}$ as the \emph{bulk} component, whereas the component with breathing frequency $\omega_{B2}$ will be referred to as the \emph{tail} component.

As we will show below---after introducing (see Sec.~\ref{Weight}) and analysing the relative weight of the bulk and tail components---the same conclusion regarding the oscillation frequencies of the bulk and tail components can be arrived at by inspecting the local dynamics of the density distribution $\rho(x,t)$ near the trap centre and in the tails of the distribution.

We pause here momentarily to emphasise the key differences between our findings and those reported in previous studies of breathing mode oscillations in a weakly interacting 1D Bose gas. While previous studies have also predicted a smooth crossover of the oscillation frequency from $\sim\!\sqrt{3}\omega$ towards $2\omega$ as the temperature is increased, the frequency in question has always been what we refer here to as the frequency of oscillations of the bulk component $\omega_{B1}$. While we observe the same crossover for the $\omega_{B1}$ component, our simulations indicate that: (\emph{i}) there is a second distinct breathing frequency $\omega_{B2}$, which is for the tail component, and (\emph{ii}) $\omega_{B2}$ undergoes a similar crossover from the value $\approx\!1.9\omega$ towards $2\omega$ as $\gamma_{0}^{3/2}\mathcal{T}$ is increased within $0.39\leq\gamma_{0}^{3/2}\mathcal{T}\leq 1$.

Apart from exciting the breathing oscillations via a sudden change of the trap frequency, we have also considered a weak sinusoidal modulation of the trap strength, $V(x,t)=[1+0.05\sin(\omega_Bt)]V(x,0)$, and of the trap frequency, $\omega(t)=[1+0.05\sin(\omega_Bt)]\omega_0$, for four oscillation periods, with three different values of the modulation frequency $\omega_B=\{\sqrt{3}\omega_0, \frac{1}{2}(\sqrt{3}\omega_0+2\omega_0),2\omega_0\}$. In these alternative protocols, we have observed no qualitative changes in the ensuing breathing oscillations. Namely, the sinusoidal modulation would still excite oscillations at two distinct frequencies displaying beating. This is again similar to the observations reported in Ref.~\cite{straatsma2016} for a partially condensed 3D Bose-Einstein condensate. Because of this, we will continue our analysis of breathing oscillations for the sudden quench protocol only, Eq.~\eqref{sudden}.

Finally, we comment on a weak dependence of our results on the choice of the cutoff energy $E_{\mathrm{cut}}$, or equivalently the maximum number of harmonic oscillator basis states $n_{\max}$ used in our simulations, with $E_{\mathrm{cut}}=\hbar\omega_0(n_{\max}+1/2)$. All results presented so far and hereafter are obtained for an optimal choice of $E_{\mathrm{cut}}$ that results from $n_{\max}=250$. As we show in Appendix~\ref{cutoff}, this choice is justified by the best match of the initial thermal equilibrium density distribution $\rho(x,t=0)$ with the density profile obtained from the solution to the exact Yang-Yang thermodynamic equations \cite{yang1969}, combined with the local density approximation \cite{kheruntsyan2005}. Finding such a match, particularly in the tails of the distribution, is a subtle problem as it is in competition with the applicability of the $c$-field approximation requiring an energy cutoff via the projector operator in Eqs.~\eqref{eq:breathing_SPGPE} and \eqref{GPE}. Moreover, the cutoff dependency depends strongly on the observable of interest that is being calculated \cite{bouchoule2016,Deuar2015,Deuar2018a,Deuar2018b,Thomas_2021,bayocboc2021dynamics}. For the breathing mode oscillations frequencies, the cutoff dependency is weak; for example, changing $n_{\max}$ from $n_{\max}=250$ to $300$ and $200$ results in less than $\pm2.2\%$ change in the extracted frequencies $\omega_{B1}$ and $\omega_{B2}$ (see Fig.~\ref{fig:frequency_diff_cutoff} in Appendix~\ref{cutoff}). Thus, while the quantitative details of our main findings have a weak cutoff dependence, the respective qualitative aspects and ensuing conclusions are essentially cutoff independent.

\subsection{Relative weight of breathing components}
\label{Weight}

To quantify the relative contribution of the beating components (bulk and tail components) to the total breathing oscillations, we introduce the relative weight of the bulk component,
\begin{equation}
	K =\frac{A_{1}^{2}}{A_{1}^{2} + A_{2}^{2}}.
\end{equation}
The relative contribution of the tail component is then given by $1-K$, and Eq.~\eqref{eq:RMS_fit} for the rms width can be rewritten as 
\begin{align}
	\label{eq:RMS_fit_K}
	\Delta x_{\mathrm{RMS}}(t) = &A \bigl[ \sqrt{K}\cos(\omega_{\mathrm{B}1}t + \phi_{1})e^{-\Gamma_{1}t} \nonumber \\
	&\,\,\,\,+ \sqrt{1-K}\cos(\omega_{\mathrm{B}2}t + \phi_{2})e^{-\Gamma_{2}t} \bigr] + C,
\end{align}
where $A = \sqrt{A_{1}^{2} + A_{2}^{2}}$.

\begin{figure}[t]
	\begin{center}
		\includegraphics[width=0.45\textwidth]{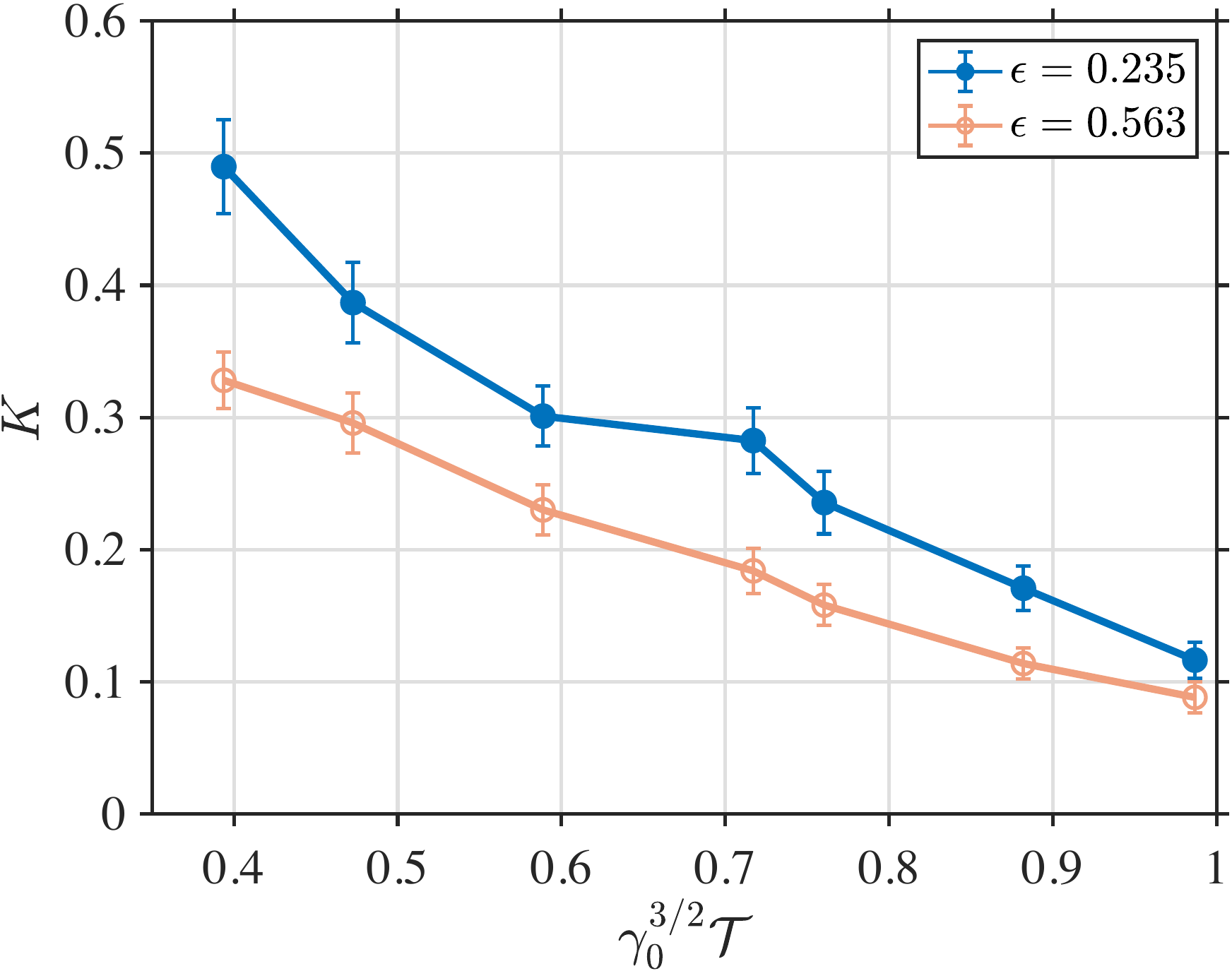}\\
		\caption{Relative weight $K$ of the bulk component in the breathing  oscillations of a 1D quasicondensate as a function of $\gamma_{0}^{3/2}\mathcal{T}$ for two different quench strengths $\epsilon$. The error bars account for the fitting error only and indicate a 95\% confidence interval. }
		\label{fig:RMS_relative_power}
	\end{center}
\end{figure}

In Fig.~\ref{fig:RMS_relative_power}, we plot the relative weight $K$ as a function of $\gamma_{0}^{3/2}\mathcal{T}$, for two different values of the quench strengths $\epsilon$ as in Fig.~\ref{fig:fit_omegaB}. As we see, the weight of the bulk component $K$, behaving predominantly as a quasicondensate with suppressed density fluctuation, but a fluctuating phase, has its maximum value for the smallest $\gamma_{0}^{3/2}\mathcal{T}$ and it decreases with $\gamma_{0}^{3/2}\mathcal{T}$. Conversely, the contribution of the tail component $1-K$, behaving as a nearly ideal degenerate Bose gas, increases with $\gamma_{0}^{3/2}\mathcal{T}$. With this observation at hand, we can now return to the results of Fig.~\ref{fig:fit_omegaB} and interpret them as follows. For system parameters deep in the quasicondensate regime ($\gamma_{0}^{3/2}\mathcal{T}\!<\!1$), a large fraction of particles occupy the low-energy states, and the bulk of the system exhibits collective breathing oscillations ($\omega_{B1}$) close to the pure mean-field behaviour of a zero-temperature system. As we go to higher values of $\gamma_{0}^{3/2}\mathcal{T}$ (by, e.g., increasing the temperature of the system, or reducing the peak density $\rho_0$ and hence increasing $\gamma_0$), a larger fraction of particles begin to thermally populate higher-energy states. As a result, a second breathing mode ($\omega_{B2}$) becomes more pronounced, with the behaviour closer to that of a degenerate ideal Bose gas. Then, as we reach the boundary with the nearly ideal degenerate Bose gas regime ($\gamma_{0}^{3/2}\mathcal{T} \sim 1$), almost all of the particles occupy high-energy modes. The collective breathing mode, characteristic of low-energy particles, begin to disappear (with $K$ going down) and the whole system now exhibits breathing oscillations with frequency $\omega_{B2}$.

\begin{figure}[tbp]
			\includegraphics[width=0.45\textwidth]{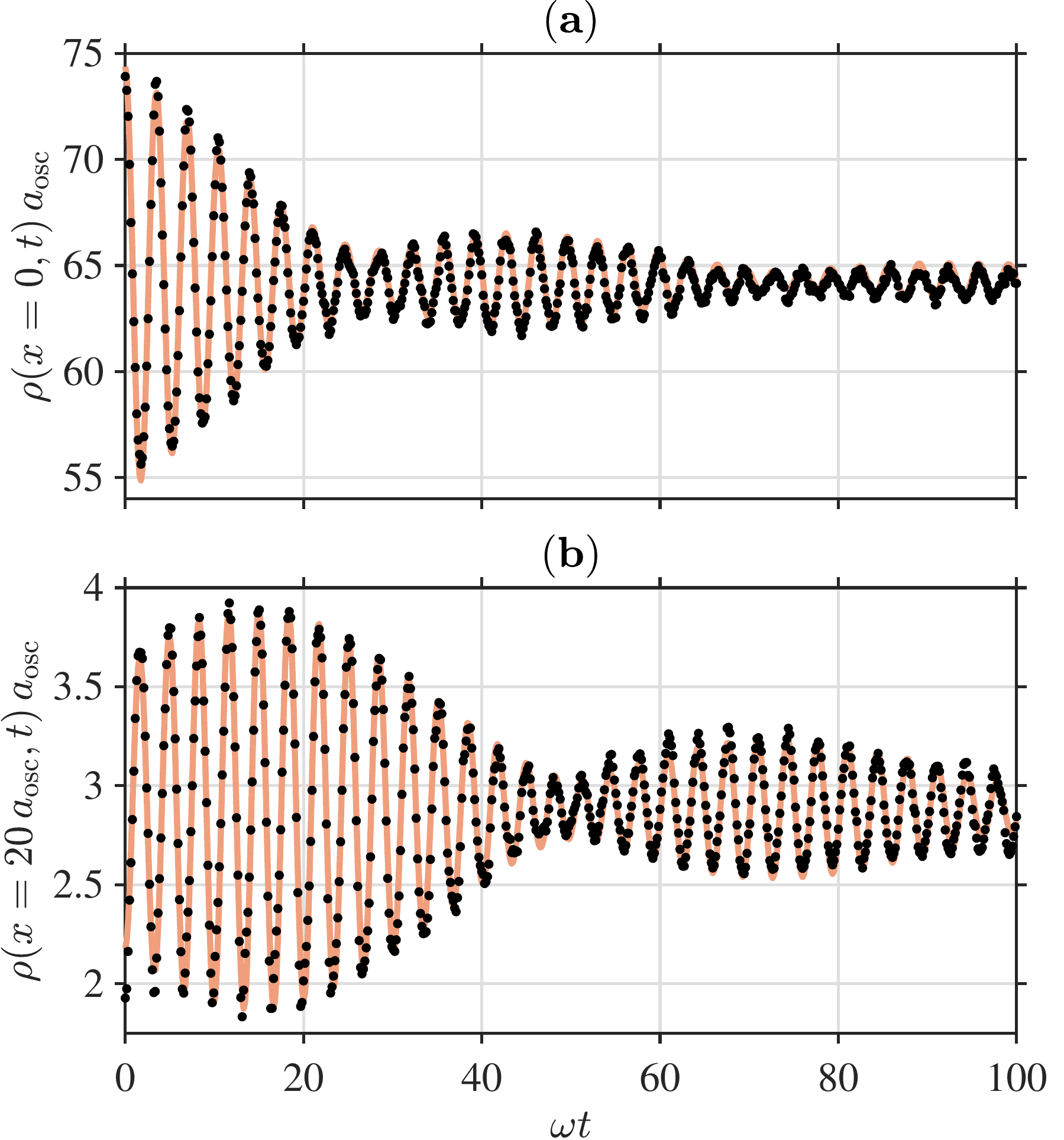}
		\caption{Local evolution of the density distribution near the trap centre, $\rho(x=0,t)$ (a), and in the tails, $\rho(x = 20\,a_{\mathrm{osc}},t)$ (b), for $\gamma_{0}^{3/2}\mathcal{T}  = 0.39$ and a quench strength of $\epsilon=0.563$ ($\omega/\omega_0 = 0.8$). 
		The black dots are data points from $c$-field simulations, whereas the full (orange) line is a fit using the right-hand-side of Eq.~\eqref{eq:RMS_fit_K}.
		 To reduce the statistical noise, we averaged the densities over a small region $x/a_{\mathrm{osc}}\in[-1,1]$ and $x/a_{\mathrm{osc}}\in[19,21]$, respectively.
		The dynamical fit to the central density $\rho(x=0,t)$ yields an oscillation frequency $\omega_{B1} = 1.77\,\omega$ with the relative weight of $K = 0.96$, and $\omega_{B2} = 1.90\,\omega$ with the respective relative weight of $1-K=0.04$. On the other hand, the fit applied to the tails, $\rho(x=20\,a_{\mathrm{osc}},t)$, yields an oscillation frequency $\omega_{B1} =1.80\,\omega$ with the relative weight of $K = 0.23$, with the other frequency being $\omega_{B2} = 1.91\omega$ and $1-K=0.77$. In both cases, the oscillation frequencies are close to those extracted from the evolution of the rms width, whereas the relative weight $K$ extracted from the tails is very different (see text).		
		}
		\label{fig:breathing_bulk_vs_tail_density}
\end{figure}

Having introduced the relative weights of the bulk ($K$) and tail ($1-K$) components, we can now also analyse the local dynamics of the density distribution $\rho(x,t)$ in the trap centre (at $x\!=\!0$) and in the tails (at $x/a_{\mathrm{osc}}\!\gg \!1$) of the distribution with the aim to further reassert that the breathing frequency $\omega_{B1}$ can be attributed to low-energy atoms populating primarily the bulk component, whereas the breathing frequency $\omega_{B2}$ can be associated with high-energy atoms populating primarily the tails of the density distribution. In Fig.~\ref{fig:breathing_bulk_vs_tail_density} we show an example of local evolution of the density distribution near the trap centre $x\!=\!0$ and in the tails around $x\! =\! 20\,a_{\mathrm{osc}}$. As we see, both curves still display beating of two frequencies; the two frequencies can be fitted with the same formula as the right-hand-side of Eq.~\eqref{eq:RMS_fit_K}; and both frequencies are approximately the same as those extracted from the rms width as in Fig.~\ref{fig:RMS_sample}. At the same time, we find that the relative weight of the $\omega_{B1}$ component in the trap centre is $K=0.96$. This is significantly larger than the value of $K$ extracted from the rms width. On the other hand, the relative weight on the same $\omega_{B1}$ component, but in the tails of the distribution, drops down to a much smaller value of $K\!=\!0.23$. Reciprocally, the relative weight ($1-K$) of the $\omega_{B2}$ component dominates the beating in the tails of the distribution, with $1\!-\!K\!=\!1\!-\!0.23\!=\!0.77$, but is much smaller in the trap centre.

\begin{figure}[tbp]
	\begin{center}
		\includegraphics[width=0.47\textwidth]{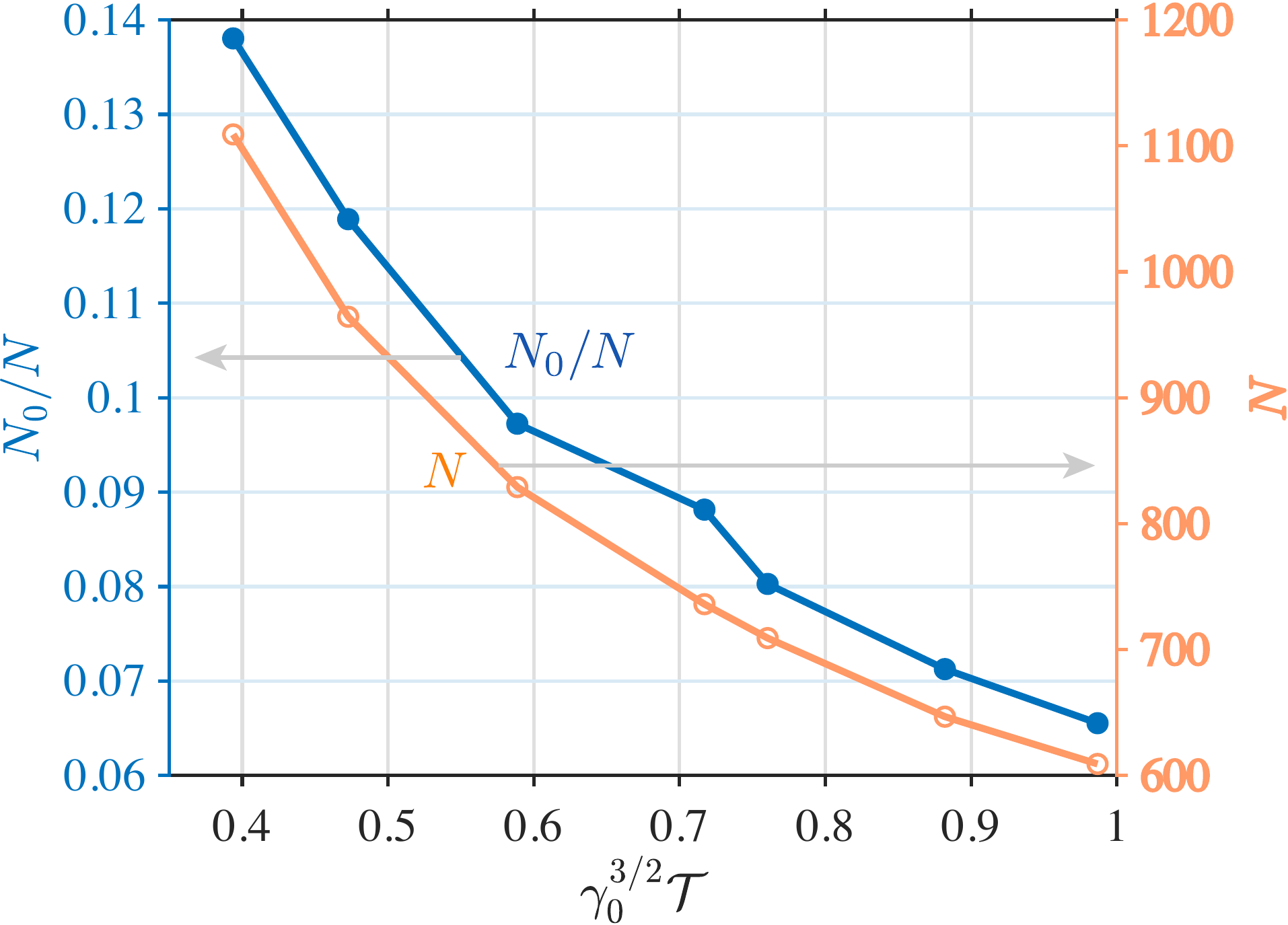}\\
		\caption{Computed condensate fraction $N_0/N$ (full circles, left vertical axis) and the respective total $N$ (open circles, right vertical axis) of the initial state of the 1D Bose gas as a function of $\gamma_{0}^{3/2}\mathcal{T}$, for a quench strength $\epsilon \simeq 0.563$. } 
		\label{fig:condensate_fraction}
	\end{center}
\end{figure}

A further insight into the composition of the bulk and tail components of the 1D Bose gas can be gained by computing the condensate fraction  $N_0/N$ of the initial state of the system as per Penrose--Onsager criterion \cite{penrose1956}. The condensate fraction $N_0/N$ is plotted in Fig.~\ref{fig:condensate_fraction} as a function of $\gamma_{0}^{3/2}\mathcal{T}$ (for a quench strength $\epsilon \simeq 0.563$), together with the respective total number of atoms in the system $N$, where we note that the dimensionless parameter $\gamma_{0}^{3/2}\mathcal{T}$ was scanned by changing the total number of particles $N$ (hence changing the peak density $\rho_0$ and $\gamma_0$) while maintaining the same absolute temperature $T$ (and hence the same value of $\mathcal{T}$).  As we see, the maximum condensate fraction, that is attained here, is approximately $0.14$ at the lowest value of $\gamma_{0}^{3/2}\mathcal{T}$, whereas the minimum condensate fraction is $\simeq\!0.07$ at the largest $\gamma_{0}^{3/2}\mathcal{T}$. For the maximum condensate fraction of only $0.14$, the corresponding relative weight $K$ of the bulk component is noticeably larger ($K\simeq\! 0.33$, from Fig.~\ref{fig:RMS_relative_power}). This implies that the bulk component is composed not only of the particles in the condensate mode, but also of particles in highly-occupied, low-energy states above the condensate mode.

\begin{figure}[tbp]
	\begin{center}
		\includegraphics[width=0.45\textwidth]{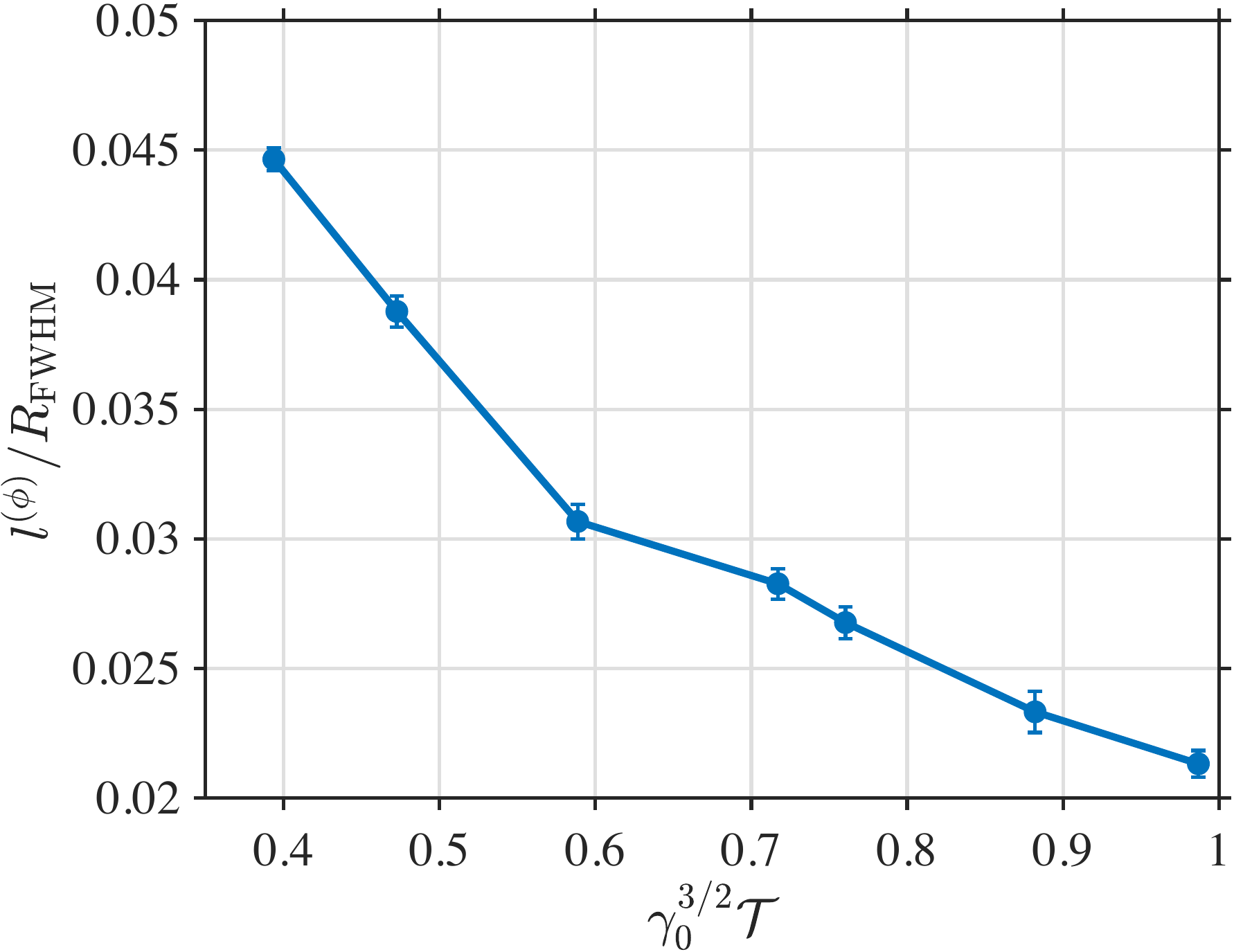}\\
		\caption{Computed thermal coherence length of the initial state of the 1D Bose gas at different values of $\gamma_{0}^{3/2}\mathcal{T}$,  for a quench strength $\epsilon \simeq 0.563$. The error bars account for the fitting error only and indicate a 95\% confidence interval.} 
		\label{fig:coherence_length}
	\end{center}
\end{figure}

The same conclusion can be arrived at by analysing an alternative quantity---the initial thermal phase coherence length $l_{\phi}$ in the trap centre---which, unlike the condensate fraction, is an intensive quantity. For a uniform quasicondensate at density $\rho$ and temperature $T$, this is given by $l_{\phi}=2\hbar^2\rho/mk_BT$ \cite{mora2003,cazalilla2004,bouchoule2012}. For a harmonically trapped system, we compute the initial ($t=0$) thermal phase coherence length in the trap centre by fitting the initial normalised first-order correlation function $g^{(1)}(x,x';t\!=\!0)=\langle \Psi^*_{\mathbf{C}}(x,0)\Psi_{\mathbf{C}}(x',0)\rangle/\sqrt{\rho(x,0)\rho(x',0)}$ at $x'=0$ with an exponential $g^{(1)}(x,x';t\!=\!0)=\exp(-|x-x'|/2l_{\phi})$ \cite{mora2003,cazalilla2004,bouchoule2012,Deuar2009}. Here, the local phase coherence length is expected to be equal to $l_{\phi}=2\hbar^2\rho_0/mk_BT$, in the local density approximation, and our fitted values are indeed very close to this analytic result. The fitted values of $l_{\phi}$ as a function $\gamma_{0}^{3/2}\mathcal{T}$ are plotted in Fig.~\ref{fig:coherence_length}, where we see qualitatively the same trend as for the condensate fraction $N_0/N$. For example, for the lowest value of $\gamma_{0}^{3/2}\mathcal{T}$ sampled in Fig.~\ref{fig:coherence_length}, the thermal phase coherence length is only a very small fraction ($\sim\!0.045$) of the full-width-at-half-maximum (FWHM) of the initial density distribution, yet the relative weight $K$ of the bulk component is $K\simeq 0.33$. This again implies that the bulk component is composed not only of the particles in the locally phase coherent region, physically similar to the condensate fraction, but extends beyond this region.

\section{Damping of breathing  oscillations}

\begin{figure}[t]
	\begin{center}
		\includegraphics[width=0.45\textwidth]{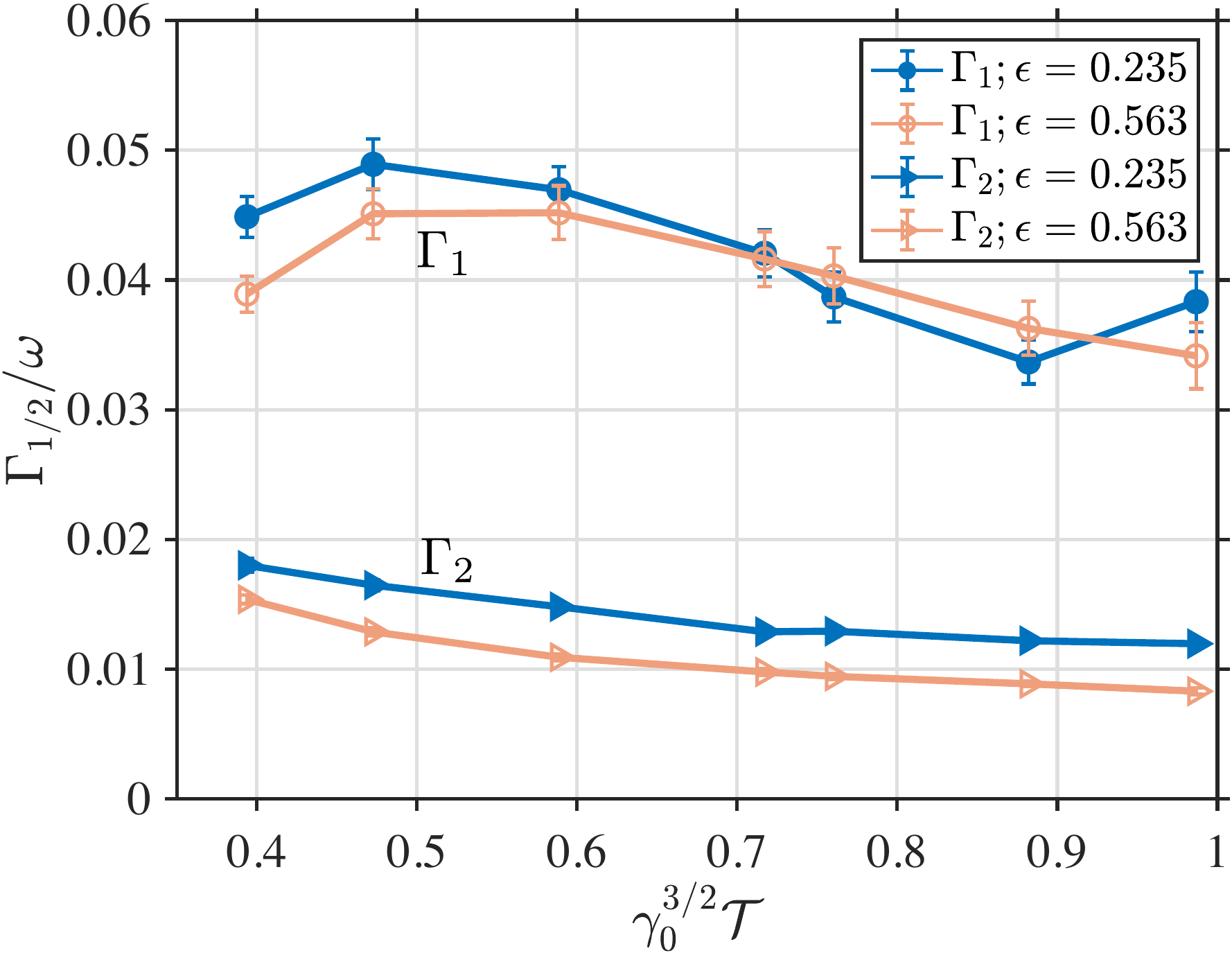}\\
		\caption{
		Damping rates $\Gamma_i$ ($i=1,2$) of the breathing oscillations in a 1D quasicondensate as a function of the dimensionless parameter $\gamma_0^{3/2}\mathcal{T}$, for two different quench strengths $\epsilon$. The error bars on data points account for the fitting error only, which indicates a 95\% confidence interval. 
		}
		\label{fig:fit_Gamma}
	\end{center}
\end{figure}

Having identified that the breathing oscillations of a 1D quasicondensate involve beating of two distinct frequencies, corresponding to the oscillations of the bulk and tail components, we now characterise the respective damping rates, $\Gamma_1$ and $\Gamma_2$, observed in Fig.~\ref{fig:RMS_sample} and extracted from fitting the results of $c$-field simulations to Eq.~\eqref{eq:RMS_fit_K}. The damping rates extracted in this way are shown in Fig.~\ref{fig:fit_Gamma} as a function of $\gamma_0^{3/2}\mathcal{T}$, for two different quench strengths $\epsilon$. Similarly to the frequencies $\omega_{\mathrm{B}1}$ and $\omega_{\mathrm{B}2}$, the damping rates $\Gamma_1$ and $\Gamma_2$ are different from each other and weakly depend on the quench strength $\epsilon$.

The damping rate $\Gamma_2$ associated with the frequency $\omega_{\mathrm{B}2}$ of the tail component is smaller than the damping rate $\Gamma_1$ associated with the frequency $\omega_{\mathrm{B}1}$ of the bulk component. This is consistent with our earlier observation that the particles comprising the tail component behave as a nearly ideal Bose gas which is expected to have very little to no damping. The damping rate $\Gamma_1$ is approximately $3-4$ times larger than $\Gamma_2$; it increases initially with $\gamma_{0}^{3/2}\mathcal{T}$, before saturating to a value of $\Gamma_{1}\! \simeq\! 0.045 \omega$ at $\gamma_{0}^{3/2}\mathcal{T}\!\simeq \!0.6$ (for a quench strength $\epsilon\!=\!0.563$) and then decreasing slightly as we approach the upper boundary of the quasicondensate regime, $\gamma_{0}^{3/2}\mathcal{T}\!\sim \!1$. For experimentally typical values of $\omega_0/2\pi\!=\! 10$ Hz (and hence $\omega/2\pi\!=\!8$ Hz for $\epsilon\!=\!0.563$), the damping rate of $\Gamma_1\!\simeq \!0.04\omega$ corresponds to $\Gamma_1\!\simeq\! 2$ s$^{-1}$ (or a damping time constant of $\tau_1\!=\!1/\Gamma_1\!\simeq \!0.5$ s), whereas the damping rate $\Gamma_2$ is four times smaller ($\tau_2\!=\!1/\Gamma_2\!\simeq \!2$ s).

Similarly to damping of low-energy collective excitations in a harmonically trapped and partially Bose-condensed 3D systems at finite temperatures ~\cite{pitaevskii1997,fedichev1998,Giorgini1998,guilleumas1999,jackson2003}, the dominant damping mechanism of the bulk component in our 1D quasicondensate can be expected to be Landau damping. In Landau damping, a low-energy collective excitation of energy $\hbar \omega_{\mathrm{B}1}$ and a thermal excitation of energy $E_i$ are annihilated (created) and another thermal excitation of energy $E_j$ is created (annihilated). Within the $c$-field approach employed in our numerical simulations, this damping mechanism is implicitly present through the interaction term in the GPE for the $c$-field $\Psi_{\mathbf{C}}(x,t)$ as the `classical region' incorporates not only the condensate mode but also many low-lying excited modes that have a relatively high thermal occupation.

Even though the damping rate of low-energy excitations due to Landau mechanism for a 1D uniform quasicondensate has been calculated in Ref.~\cite{yang2012}, we defer a critical analysis of this work and its possible application to our trapped system to a future study. We nevertheless note here that any possible comparison of the damping rate $\Gamma_1$ from $c$-field simulations and that obtained from Landau mechanism should take into account the fact that in the $c$-field approximation the `classical region' includes only a fraction of thermal particles (that belong to highly occupied modes) and hence the extracted value of $\Gamma_1$ is likely to underestimate the damping rate compared to alternative predictions that take into account all thermal particles. At the same time, the theory of Landau damping conventionally assumes that the thermal excitations are always in thermal equilibrium, whereas this assumption does not apply to our system because quenching the trap frequency also excited a collective breathing oscillation of the tail component, which acts as a dynamical (rather than a static) bath of thermal excitations.

In an equivalent quench scenario in a 3D system, these questions can, in principle, be addressed using, e.g., the Zaremba-Nikuni-Griffin (ZNG) formalism \cite{zaremba1999,griffin_nikuni_zaremba_2009}, where the condensate part of the system is described by the generalised Gross-Pitaevskii equation, whereas the noncondensate (thermal) part is described by the quantum Boltzmann equation. However, the ZNG formalism cannot be directly applied to 1D quasicondensate due the fact that the fractional occupancy of the ground-state condensate mode in 1D does not dominate the occupancies of excited modes as it does in 3D. Accordingly, a simple separation into a condensate and thermal excitations is not justified here. For the same reasons, our observation of two excitation modes cannot be simply interpreted as a consequence of a standard two-fluid model applicable to 3D systems, where the two fluids are represented by the superfluid (condensate) and the normal (thermal) components of the gas and where such a phenomenon would be common \cite{jin1997,straatsma2016}. Apart from this, we note that the two-fluid model is applicable in the strongly collisional regime (when the damping rate is much larger than the angular frequency of excitation modes), whereas the breathing mode excitations studied in this work are in the opposite weakly collisional regime. Overall, our finding call for a further study beyond the classical field approximation, which could perhaps be accomplished by generalising the GNZ formalism to 1D systems, wherein the evolution of the entire $c$-field would be coupled to a quantum Boltzmann equation.

\section{Summary}

In conclusion, we have studied the breathing oscillations of a harmonically trapped 1D Bose gas in the quasicondensate regime, invoked after a sudden quench of the trap frequency. Using the $c$-field approach for sampling the initial thermal equilibrium state and simulating the subsequent post-quench dynamics, we observed beating of two breathing modes. The two breathing modes oscillate at two distinct frequencies $\omega_{Bi}$ ($i\!=\!1,2$) and have their own damping rates $\Gamma_{i}$. 
Furthermore, they can respectively be attributed to low-energy particles in the bulk and high-energy particles in the tails of the density distribution of the gas. The bulk component breathes with the frequency close to the expected breathing mode frequency of a zero temperature system, $\omega_{B1}\simeq \sqrt{3}\omega$ for most of the quasicondensate region, whereas the breathing mode frequency of the tail component is closer to that of an ideal Bose gas, $\omega_{B2}\simeq 2\omega$.  The damping rates $\Gamma_1$ and $\Gamma_2$, extracted from the $c$-field simulations for typical experimental parameters, have the associated damping time constants on the order of $0.5$\,s and $2$\,s, respectively, for most of the values of $\gamma_0^{3/2}\mathcal{T}$ considered.

In order to experimentally observe the predicted beating of two breathing modes, one needs to ensure that the breathing dynamics is monitored for sufficiently long time as to detect reduction and subsequent revival of the amplitude of oscillations due to beating. One has to also ensure that the system is deep in the 1D regime as to eliminate additional damping mechanisms due to transverse excitations, which can prevent the revivals. Taking the beat frequency $\omega_{\mathrm{beat}}=\omega_{B2}-\omega_{B1}$ as simply $\omega_{\mathrm{beat}}=2\omega-\sqrt{3}\omega$, and assuming $\omega/2\pi=8$ Hz, one obtains the beat period of the order of $0.5$ seconds. This is well within the reach of the current 1D Bose gas experiments.

\emph{Note added.} After completing this work and submitting the manuscript, we became aware of Ref.~\cite{Robertson:2022} which analyses phonon decay in 1D quasicondensates via the Landau-Beliaev damping mechanism. The analysis of Ref.~\cite{Robertson:2022} goes beyond that of Ref.~\cite{yang2012} in terms of underlying assumptions and considerations, however, their analytic prediction for the damping rate appears to be applicable to a different regime than the low-energy breathing excitations studied in the present work. Accordingly, a direct comparison of our numerical results with the said analytic prediction is not possible at the moment. This in turn highlights the need for either revisiting and extending the theory of Landau-Beliaev damping of trapped 1D Bose gases, or else developing an analytic understanding of the decay of 1D breathing oscillations via an alternative mechanism.

\begin{acknowledgments}
F.\,A.\,B. and K.\,V.\,K. acknowledge stimulating discussions with I. Bouchoule and M. J. Davis.  K.\,V.\,K. acknowledges support by the Australian Research 
 Council Discovery Project Grants DP170101423 and DP190101515. 
 \end{acknowledgments}

\appendix

\section{Regimes of a weakly interacting uniform 1D Bose gas}
\label{appendix:Regimes}

\begin{figure}[tbp]
		\includegraphics[width=0.97\linewidth]{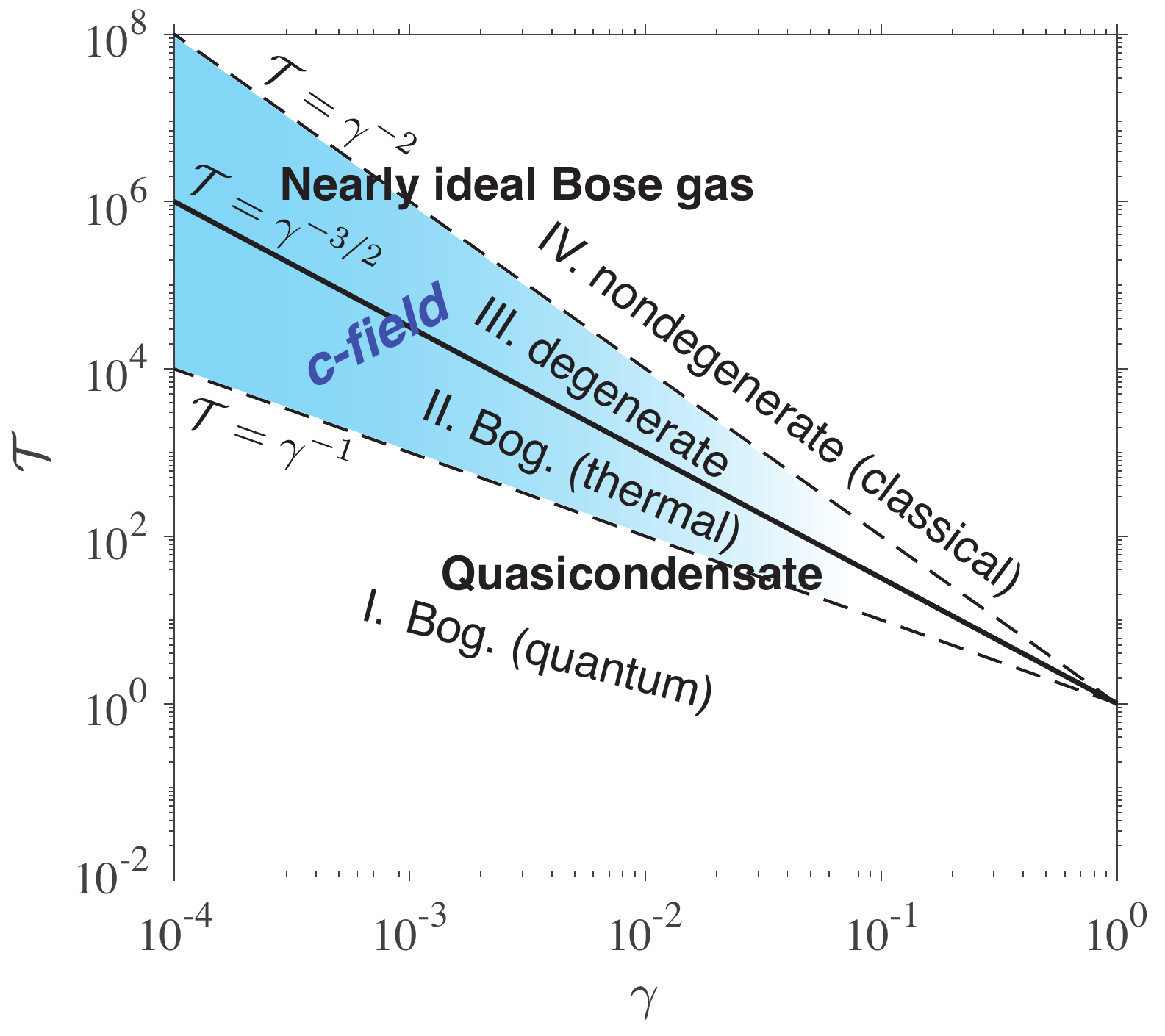}
		\caption{
		Crossover phase diagram of a weakly interacting ($\gamma\ll 1$) uniform 1D Bose gas, showing the classification of different sub-regimes in the parameter space $(\gamma,\mathcal{T})$, and the region of applicability of the $c$-field approach (see text). }		
		\label{fig:phase_diagram}
\end{figure}

For a uniform 1D Bose gas at linear density $\rho$, the different thermal equilibrium regimes have been identified in Refs.~\cite{kheruntsyan2003,kheruntsyan2005} through the analysis of local density-density correlation function. They 
can be characterised by just two parameters, the dimensionless interaction strength $\gamma \!=\! mg/\hbar^{2}\rho_0$ and dimensionless temperature $\mathcal{T} \!=\! k_{\mathrm{B}}T/(mg^{2}/2\hbar^2)$. An alternative choice for the dimensionless temperature is to define it via the temperature of quantum degeneracy, $T_d=\hbar^2\rho^2/2m$, via $\tau=k_{\mathrm{B}}T/T_d$ \cite{kheruntsyan2003}, where we note that the two temperatures are related by $\tau=\mathcal{T}\gamma^2$. However, as was shown in Ref~\cite{kheruntsyan2005}, the temperature $\mathcal{T}$ (unlike $\tau$) is more convenient to also characterise an inhomogeneous system within the local density approximation as it can serve as the global temperature of the system that does not depend on the local density. The phase diagram in the $(\gamma,\mathcal{T})$ parameter space \cite{kheruntsyan2003}, for a weakly interacting system, $\gamma\ll 1$, is shown in Fig.~\ref{fig:phase_diagram}, where we note that the different sub-regimes are smooth crossovers. In this diagram, the different sub-regimes that can be treated analytically using approximate theoretical approaches are first introduced at the level of a quasicondensate, characterised by suppressed density fluctuations (similar to a true condensate) but fluctuating phase \cite{petrov2000}, and a nearly ideal Bose gas in which both the density and phase fluctuate.

In terms of the underlying theoretical approaches, the quasicondensate regime can be described by the Bogoliubov theory \cite{kheruntsyan2003,Sykes2009,Deuar2009}, in which one can further distinguish the sub-regimes dominated by quantum (region I) or thermal (region II) fluctuations, corresponding, respectively to $k_{\mathrm{B}}T\ll g\rho$ (or $\mathcal{T}\ll \gamma^{-1}$) and $g\rho\ll k_{\mathrm{B}}T\ll \sqrt{\gamma}\hbar^2\rho^2/2m$ (or $\gamma^{-1}\ll \mathcal{T}\ll\gamma^{-3/2}$). In Ref.~\cite{kheruntsyan2003}, these sub-regimes were referred to as GPa and GPb regimes, respectively.

The nearly ideal Bose gas, on the other hand, can be treated using the perturbation theory with respect to $\gamma$ around the ideal (noninteracting) Bose gas \cite{kheruntsyan2003,Sykes2009,Deuar2009}. Here, we can further distinguish between a highly \emph{degenerate} nearly ideal Bose gas ($\tau\!\ll \!1$, region III) and a \emph{nondegenerate}, nearly \emph{classical} ideal gas ($\tau\!\gg\!1$, region IV). These two sub-regimes, correspond, respectively, to $\sqrt{\gamma}\hbar^2\rho^2/2m \ll k_{\mathrm{B}}T\ll \hbar^2\rho^2/2m$ (or $\gamma^{-3/2}\ll \mathcal{T}\ll \gamma^{-2}$) and $\hbar^2\rho^2/2m \ll k_{\mathrm{B}}T$ (or $\gamma^{-2}\ll \mathcal{T}$). In Ref.~\cite{kheruntsyan2003}, these sub-regimes were referred to as ``decoherent quantum'' (DQ)  and ``decoherent classical'' (DC) regimes, respectively.

\section{Regime of applicability of the $c$-field approach}
\label{c-field-regime}

Turning now to the region of applicability of the $c$-field approach, we note that it is adequate for describing thermal (rather than quantum) fluctuations in highly degenerate Bose gases. As such, in the phase diagram of Fig.~\ref{fig:phase_diagram}, it spans the sub-regimes II and III. 

More explicitly, the condition of applicability of the $c$-field approach, $|\mu| \ll k_{\mathrm{B}}T$ \cite{castin2000}, where $\mu$ is the chemical potential, can be rewritten as $g\rho \ll k_{\mathrm{B}}T$ in the thermal quasicondensate sub-regime II, where $\mu\simeq g\rho$. In dimensionless from, this coincides with the temperature lower bound on the thermal quasicondensate regime, $\gamma^{-1}\ll \mathcal{T}$. 

On the other hand, in the degenerate ideal Bose gas sub-regime III, the absolute value of the chemical potential ($\mu<0$) can be approximated by $|\mu|\simeq m(k_BT)^2/2\hbar^2\rho^2$ \cite{castin2000,Bouchoule2007}. Hence, the condition of applicability of the $c$-field approach here can be rewritten as $k_BT\ll 2\hbar^2\rho^2/m$, which we note agrees (ignoring numerical factors of the order of one) with the upper bound on the dimensionless temperature $\mathcal{T}$ in the degenerate nearly ideal Bose gas regime, $\mathcal{T}\ll\gamma^{-2}$. 
Combining the two sub-regions, $\gamma^{-1}\ll \mathcal{T}$ and $\mathcal{T}\ll\gamma^{-2}$, gives $\gamma^{-1}\ll \mathcal{T} \ll \gamma^{-2}$ for the regime of applicability of the $c$-field approach, which is shown as the shaded area in Fig.~\ref{fig:phase_diagram}.

Furthermore, a remarkable property of the $c$-field approach is that one can show (after introducing appropriately defined time-, length-, and energy-scales; see Refs.~\cite{castin2000,bouchoule2012,bouchoule2016,Thomas_2021} for details) 
that the corresponding equations of motion can be rewritten in a dimensionless form in such a way that they depend only on a single dimensionless parameter $\gamma^{3/2}\mathcal{T}$ (with $\gamma^{3/2}\mathcal{T}\equiv 2k_BT/\sqrt{g\rho(\hbar^2\rho^2/m)}$, rather than on two independent dimensionless parameters $\gamma$ and $\mathcal{T}$. In terms of this single parameter, the two relevant regimes of the weakly interacting 1D Bose gas 
can be rewritten as $ \gamma^{1/2}\ll \gamma^{3/2}\mathcal{T} \ll 1$ (region II), and $1 \ll \gamma^{3/2}\mathcal{T} \ll \gamma^{-1/2}$ (region III), whereas the overall region of applicability of the $c$-field approach is obtained by combining the two,
 $\gamma^{1/2}\ll \gamma^{3/2}\mathcal{T} \ll \gamma^{-1/2}$. 

As a further remark, we note here that the crossover boundaries between the different regimes of a weakly interacting 1D Bose gas, that are dominated by thermal rather than quantum fluctuations, were identified here through the properties of short-range density-density or second-order correlation functions. If, however, one is concerned with the behaviour of the first-order or phase correlation function at large relative distances, or equivalently the momentum distribution at low momenta, then the lower bound on the temperature, in which the physics is dominated by thermal fluctuations, is reduced from $g\rho\ll k_{\mathrm{B}}T$ down to $g\rho e^{-2\pi/\sqrt{\gamma}}\ll k_BT$; for further details, see footnotes [59] and [63] of Ref.~\cite{bouchoule2012}.

\section{Regimes of a harmonically trapped gas}
\label{trapped}

In a harmonically trapped (inhomogeneous) 1D Bose gas, with longitudinal trap potential $V(x)=\frac{1}{2}m\omega^2x^2$, the linear density $\rho$ becomes position-dependent and describes the density profile of the gas $\rho(x)$. Accordingly, the dimensionless interaction strength also becomes position dependent, $\gamma(x)= m g/(\hbar^2 \rho(x))$, while the dimensionless temperature ${\mathcal{T}}$ continues to serve as a global equilibrium temperature of the system. For a given chemical potential $\mu$, which fixes the total number of particles $N$ in the system in a given trap, the density profile $\rho(x)$ and its peak value in the trap centre $\rho_0\equiv \rho(0)$ are unique. Therefore the peak density $\rho_0$ can be used to define a dimensionless interaction strength $\gamma_0=m g/(\hbar^2 \rho_0)$ that plays the role of a global interaction parameter for the entire system.

As shown in Refs.~\cite{bouchoule2016,Thomas_2021}, the combination $\gamma_0^{3/2}\mathcal{T}$ remains relevant for efficient parametrisation of a harmonically trapped system, except that now one needs an additional parameter---the trap frequency $\omega$---to completely characterise the system. The dimensionless trap frequency $\overline{\omega}$ is defined according to $\bar{\omega}= \omega \hbar^{5/3}m^{1/2}/[g^{2/3}(k_BT)^{2/3}]$, and we refer the reader to Ref.~\cite{Thomas_2021} for further details.

\begin{figure*}[tbp]
\includegraphics[width=14cm]{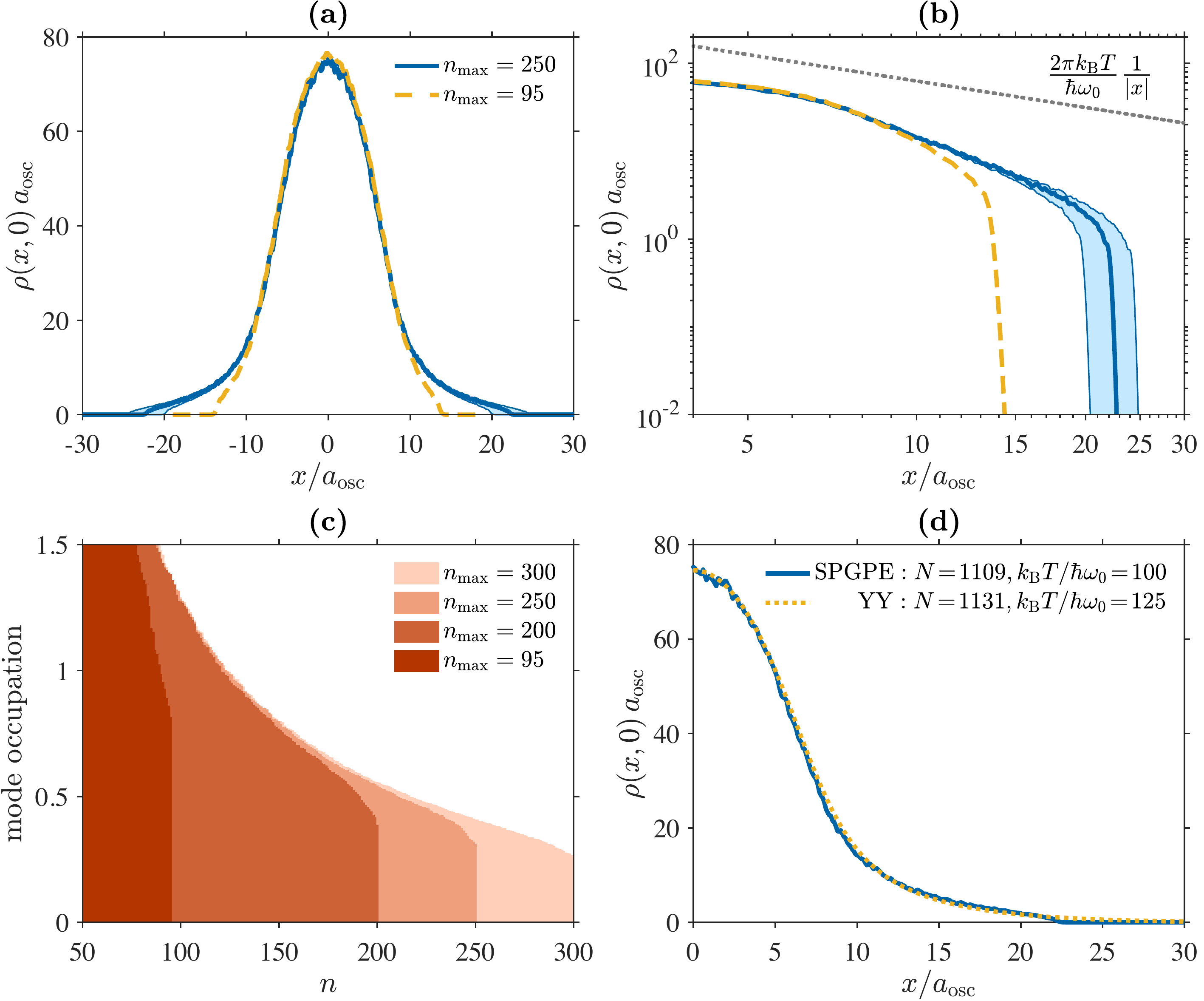}
	\caption{Density profiles and mode occupancies of the initial thermal equilibrium quasicondensate for $\gamma_0^{3/2}\mathcal{T}= 0.39$ ($N=1109$) corresponding to the first data point in Fig.~\ref{fig:fit_omegaB} of the main text. The thick solid (blue) line in (a) and (b) corresponds to the cutoff choice of $n_{\max}=250$ in the SPGPE simulations, whereas the upper and lower bounds of the shaded areas around this curve represent, respectively, the density profiles for $n_{\max}=300$ and $n_{\max}=200$. The dashed (yellow) line corresponds to the SPGPE density profile calculated with the cutoff of $n_{\max}=95$. The straight dotted line in (b) shows the $\propto\!1/|x|$ scaling of the tails of the density profile expected for an ideal (noninteracting) Bose gas under mode occupancies of $\gg1$.  Panel (c) shows the actual occupancies of high energy modes, for different values of $n_{\max}$. In panel (d), we show the SPGPE density profile (in the region $x\geq0$) for $n_{\max}=250$ and a best-fit density profile obtained using Yang-Yang thermodynamics within the local density approximation. We note that any such fitting is sensitive to the values of the chemical potential in the trap centre $\mu_0$, or alternatively the total number of atoms $N$, and the temperature $T$ of the system: due to the high-energy cutoff imposed in the SPGPE approach, matching up the chemical potentials and temperatures in order to obtain the same peak densities within the two methods leads to a noticeable mismatch in $N$ because the $c$-field approach cuts off the tails of the density profile. Using instead the total atom number $N$ and temperature $T$ as the two fitting parameters (for a given trap frequency), and noting that both methods are approximate, we obtain a much better overall agreement between the two profiles. This can be seen in panel (d), where the optimal parameters correspond to a slightly higher Yang-Yang temperature (as shown in the figure legend), but the corresponding total atom numbers are closer to each other compared to the case of matching up the chemical potentials in the trap centre.	}
	\label{fig:density_largest_gas}
\end{figure*}

The remaining consideration that needs to be taken into account here---in view of the region of applicability of the $c$-field approach to a harmonically trapped system---is the following. If the temperature and the peak density of the trapped gas is chosen to lie in the thermal quasicondensate region II, $\gamma_0^{1/2}\ll\gamma_0^{3/2}\mathcal{T} \!<\!1$, then, for typical experimental parameters, such as $\gamma_0\sim 10^{-3}-10^{-2}$ and $\mathcal{T}\sim 10^2-10^3$, the tails of the density distribution will end up lying in the nearly ideal degenerate Bose gas regime III, $1< \gamma(x)^{3/2}\mathcal{T}\ll\gamma(x)^{-1/2}$. This is because the local density $\rho(x)$ decreases with $x$ away from the trap centre $x\!=\!0$, leading to increasing values of $\gamma(x)$. Accordingly, the local conditions of the gas (in the local density approximation sense, with the local chemical potential given by $\mu(x)=\mu-V(x)$ \cite{kheruntsyan2005}) will---at some position $x$---cross the upper bound of region II and enter the region III. This is not an issue for the $c$-field approach, because it is applicable in both regions II and III. If, on the other hand, the peak density already lies in region III, then the tails of the distribution may end up (depending on the value of $\mathcal{T}$) in the nearly classical ideal gas regime IV, where the $c$-field approach is no longer applicable. For these reasons, all numerical simulations carried out in this work are restricted to the range of $\gamma_0^{3/2}\mathcal{T}\in[0.39,1]$ (with $\mathcal{T}\simeq 517$ kept the same), so that both the bulk and the tails of the gas lie in the regime of applicability of the $c$-field approach.

\section{Cutoff dependencies}
\label{cutoff}

As a rule of thumb, the high-energy cutoff $E_{\mathrm{cut}}$ in the $c$-field approach is usually chosen in such a way that the mode occupancy $\overline{N}_{\mathrm{cut}}$ of the highest energy mode included in the computational basis remains larger than or on the order of one \cite{castin2000,blakie2008}. Even then, the optimum cutoff depends strongly on the observable of interest that is being calculated and hence can somewhat deviate form this common prescription. Furthermore, as was recently shown in a series of works by Pietraszewicz and Deuar \cite{Deuar2015,Deuar2018a,Deuar2018b}, the optimum cutoff in energy $E_{\mathrm{cut}}$ for 1D Bose gases should be chosen even higher than in most prior determinations; in terms of the cutoff mode occupancy, it corresponds to $\overline{N}_{\mathrm{cut}}\simeq 0.78$, which is lower than one. As shown in Refs.~\cite{Deuar2015,Deuar2018a,Deuar2018b}, this lower optimum value of $\overline{N}_{\mathrm{cut}}$ is needed to obtain the correct kinetic energy of the 1D Bose gas, but does not detrimentally affect other observables.

In this appendix, we go somewhat further than this and show that even a lower cutoff mode occupancy is required in our simulations in order to faithfully reproduce the tails of the density distributions, so that they are closer to the density profiles that can be obtained from exact Yang-Yang thermodynamics in the local density approximation \cite{kheruntsyan2005}. As such, the simulations in the main text are performed with a cutoff mode occupancy of $\overline{N}_{\mathrm{cut}}\simeq 0.31$. This cutoff itself originates from using a maximum number of $n_{\max}=250$ harmonic oscillator basis modes (Hermite-Gauss polynomials). This value of $\overline{N}_{\mathrm{cut}}$ for $n_{\max}=250$ is obtained for the smallest value of the dimensionless parameter $\gamma_0^{3/2}\mathcal{T}\simeq 0.39$, which in turn corresponds to the largest quasicondensate considered (with $N=1109$). For all larger values of $\gamma_0^{3/2}\mathcal{T}$, the cutoff mode occupancy resulting from the same $n_{\max}=250$ remains essentially the same, reaching the value $\overline{N}_{\mathrm{cut}}\simeq 0.32$ for $\gamma_0^{3/2}\mathcal{T}\simeq 1$ corresponding to the smallest quasicondensate considered (with $N=609$). 

We additionally emphasise that the use of Hermite-Gauss polynomials is known as the most computationally efficient and optimal basis choice for implementing the SPGPE and projected GPE simulations \cite{blakie2008,Blakie_2008_Numerical_Method,rooney2014}, compared to the more commonly used plane-wave basis where additional subtleties arise regarding the optimal choice of the cutoff in momentum space \cite{Bradley_2005}.

In Fig.~\ref{fig:density_largest_gas} we show the density profiles of the initial thermal equilibrium quasicondensate, for the smallest value of $\gamma_0^{3/2}\mathcal{T}= 0.39$ (largest $N=1109$), corresponding to the first data point in Fig.~\ref{fig:fit_omegaB} of the main text. The different profiles are evaluated for four different values of the cutoff energy $E_{\mathrm{cut}}$, corresponding to $n_{\max}=95,200,250,300$. Figure~\ref{fig:density_largest_gas}\,(a) shows the full density profiles, whereas Fig.~\ref{fig:density_largest_gas}\,(b) zooms into the tails of the distributions. While the values of $n_{\max}=200$ and $300$ are chosen to illustrate small variations (weak cutoff dependence) of the density profiles around the case with $n_{\max}=250$, the lowest value of $n_{\max}=95$ is chosen in such a way that it results in the cutoff mode occupancy of $\overline{N}_{\mathrm{cut}}\simeq 0.78$ prescribed in Refs.~\cite{Deuar2015,Deuar2018a,Deuar2018b}.

\begin{figure}[tbp]
	\includegraphics[width=0.45\textwidth]{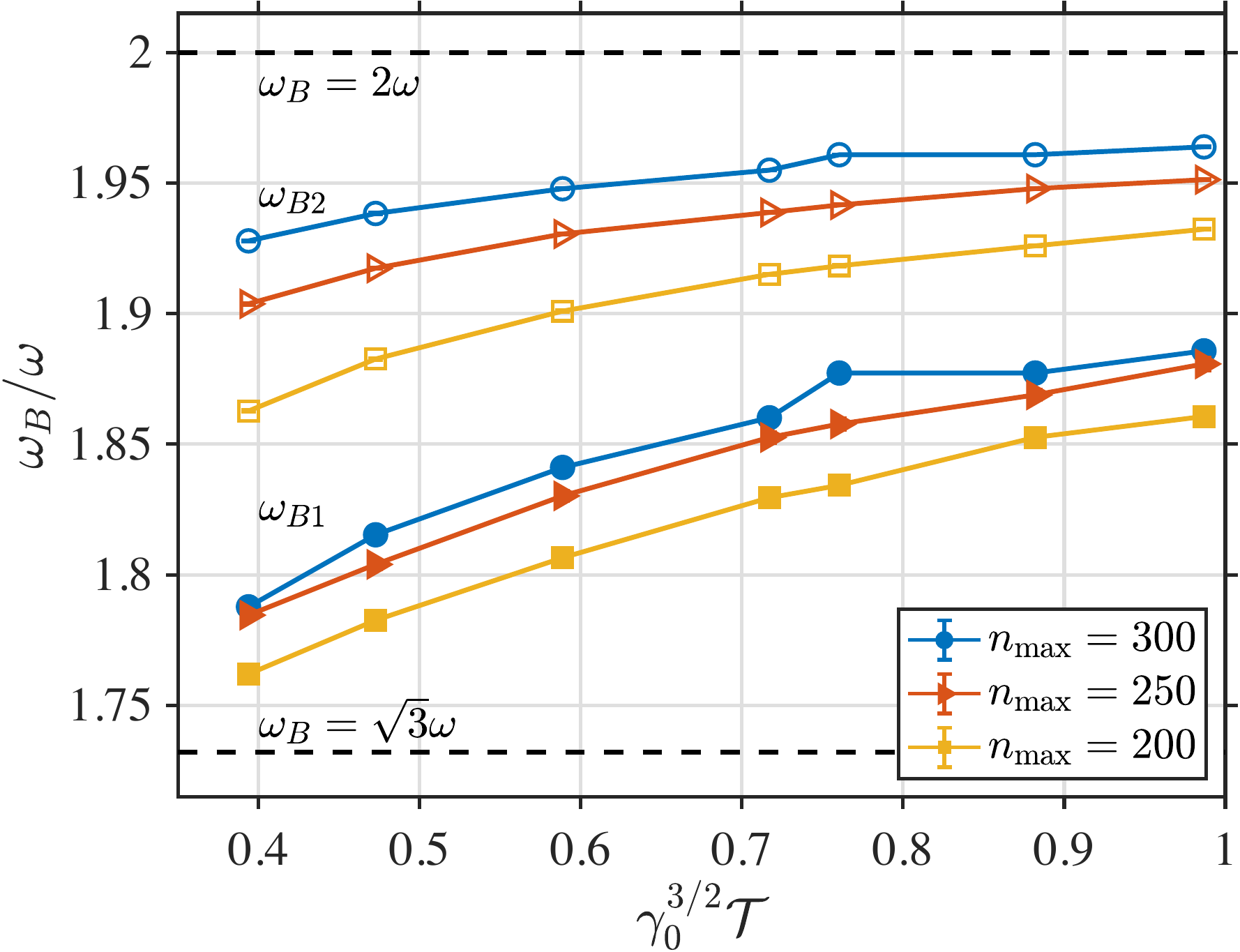}
	\caption{Same as in Fig.~\ref{fig:fit_omegaB} of the main text for $\epsilon \simeq 0.563$ and $n_{\max}=250$ (triangles), except that we also show the extracted data for simulations with $n_{\max}=200$ (squares) and $n_{\max}=300$ (circles). The error bars of data points are on the same order of magnitude as in Fig.~\ref{fig:fit_omegaB} and are no longer shown.   	}
	\label{fig:frequency_diff_cutoff}
\end{figure}

The actual mode occupancies for all four cases are shown in Fig.~\ref{fig:density_largest_gas}\,(c) and are evaluated as in Ref.~\cite{Blakie_2008_Numerical_Method}, using diagonalisation of the reduced one-body density matrix in harmonic oscillator basis. In the diagonalised density matrix, the cutoff mode occupancy (i.e., the occupancy of the highest energy mode at $E_{\mathrm{cut}}$) corresponds to the smallest nonzero eigenvalue. As we see, for all three near-optimum cases of $n_{\max}=200,250,300$, the cutoff mode occupancy $\overline{N}_{\mathrm{cut}}$ varies between $\overline{N}_{\mathrm{cut}}=0.25-0.40$. For all these cutoffs, the SPGPE density profiles can be matched up well with the density profiles calculated using the Yang-Yang thermodynamics; an example of a profile is shown in Fig.~\ref{fig:density_largest_gas}\,(c) for $n_{\max}=250$. On the other hand, for $n_{\max}\simeq95$ (with $\overline{N}_{\mathrm{cut}}\simeq 0.78$) the density profile lacks any tails, which is physically unrealistic, and it cannot be fitted well with any Yang-Yang profile. These results justify our choice of $n_{\max}\simeq250$ as an optimal cutoff for all simulation results discussed in the main text.

Furthermore, in Fig.~\ref{fig:density_largest_gas}\,(b) we show $1/|x|$ scaling of the tails of the density profile, expected under the strict $c$-field approximation of large ($\gg\!1$) occupancies of all modes, including the cutoff mode occupancy. Here, in the low-density part of the density profile, the particles can be treated locally as in an ideal (noninteracting) Bose gas, with the semiclassical distribution in position-momentum phase space given by $\rho(k,x)\simeq k_\mathrm{B}T/[\hbar^2k^2/2m+m\omega_0^2x^2/2]$. Integrating $\rho(k,x)$ over $k$ gives $\rho_\mathrm{tails}(x)\simeq 2\pi k_{\mathrm{B}}T/\hbar\omega_0|x|$, which is shown as a straight (in logarithmic scale) dotted line in Fig.~\ref{fig:density_largest_gas}\,(b). As we see, the actual tails of the density profiles calculated using the $c$-field SPGPE approach with $n_{\max}=\{200, 250, 300\}$ decay faster than the $1/|x|$ scaling expected from the `orthodox' $c$-field approximation. (The faster-than $1/|x|$ decay is a result of explicit use of the projector operator in Eq.~\eqref{eq:breathing_SPGPE}, which we recall is automatically implemented in harmonic oscillator basis by the finite number of $n_{\max}$ of modes used in the simulations.) This again justifies our somewhat unorthodox choice of $n_{\max}$ and the corresponding relatively low cutoff mode occupancy: by allowing for occupancies smaller than one, we are able to faithfully represent the tails of the density distribution by incorporating regions with sufficiently large $|x|$. Yet, we are not penalised by reaching the unphysical $1/|x|$ scaling, under which the integrated density profile or its rms width would \emph{a priory} diverge. Instead, the rms width and the extracted breathing oscillation frequencies obtained from the SPGPE with a `hard', projector-imposed cutoff show only weak cutoff dependence in the optimal range of $n_{\max}=200-300$. This can indeed be seen in Fig.~\ref{fig:frequency_diff_cutoff}, where we show the breathing oscillation frequencies as in Fig.~\ref{fig:fit_omegaB} obtained with $n_{\max}=250$ (for the case $\epsilon=0.563$), except that we additionally include the data extracted from simulations with $n_{\max}=200$ and $n_{\max}=300$. As we see, the cutoff dependence is relatively weak: changing $n_{\max}$ from $250$ to $200$ or $300$ results in less than $\pm2.3\%$ difference in the extracted values of  $\omega_{B1}$ and $\omega_{B2}$ (with both curves shifting up or down as the shift in $n_{\max}$) compared to the values extracted for $n_{\max}=250$.

As a final remark, we mention that all of the observations discussed above with regard to Fig.~\ref{fig:density_largest_gas} remain qualitatively unchanged when we scan the dimensionless parameter $\gamma_0^{3/2}\mathcal{T}$ from its smallest to the largest value of $\gamma_0^{3/2}\mathcal{T}\simeq1$, corresponding to the last data point in Fig.~\ref{fig:fit_omegaB} of the main text, or the smallest quasicondensate considered, with $N=609$.


%

\end{document}